\documentclass[aip,jcp,citeautoscript,nofootinbib,reprint]{revtex4-1}
\usepackage[utf8]{inputenc}
\usepackage{color}
\usepackage{graphicx}% Include figure files
\usepackage{dcolumn}% Align table columns on decimal point
\usepackage{bm}% bold math
\usepackage{epsfig}
\usepackage{graphicx}
\usepackage{epstopdf}
\usepackage{amsmath}
\usepackage{amssymb}
\usepackage{braket}
\newcommand{\Tr}{\text{Tr}}

\newcommand{\pnot}{\mathbf{p}_0}

\newcommand{\pnotp}{\mathbf{p}^\prime_0}
\newcommand{\cq}{\mathbf{c}_{\mathbf{q}}}
\newcommand{\cpe}{\mathbf{c}_{\bm{\rho}}}

\newcommand{\DPnot}{\Delta_{P_0}}
\newcommand{\DRnot}{\Delta_{R_0}}

\newcommand{\mqqf}{\mathbf{M}_{qq}^f}
\newcommand{\mqpf}{\mathbf{M}_{q\rho}^f}
\newcommand{\mpqf}{\mathbf{M}_{\rho q}^f}
\newcommand{\mppf}{\mathbf{M}_{\rho\rho}^f}
\newcommand{\mqqb}{\mathbf{M}_{qq}^b}
\newcommand{\mqpb}{\mathbf{M}_{q\rho}^b}
\newcommand{\mpqb}{\mathbf{M}_{\rho q}^b}
\newcommand{\mppb}{\mathbf{M}_{\rho\rho}^b}

\newcommand{\gamz}{\bm{\gamma}_0}
\newcommand{\gamt}{\bm{\gamma}_t}
\newcommand{\G}{\mathbf{G}}

\newcommand{\ptil}{\mathbf{\tilde{p}}}
\newcommand{\qtil}{\mathbf{\tilde{x}}}

\newcommand{\dx}{\int\text{d}}

\newcommand{\bz}{\mathbf{z}}
\newcommand{\opA}{\hat{A}}
\newcommand{\opB}{\hat{B}}
\newcommand{\opH}{\hat{H}}

\newcommand{\znot}{\mathbf{z}_0}
\newcommand{\znotp}{\mathbf{z}_0^\prime}

\newcommand{\zt}{\mathbf{z}_t}
\newcommand{\ztp}{\mathbf{z}_t^\prime}
\newcommand{\bigP}{\mathbf{P}}
\newcommand{\bigR}{\mathbf{R}}
\newcommand{\pe}{\mathbf{p}}

\newcommand{\xe}{\mathbf{x}}
\newcommand{\bigV}{\mathbf{V}}
\newcommand{\bigS}{\mathbf{S}}
\newcommand{\bigC}{\mathbf{C}}
\newcommand{\bigD}{\mathbf{D}}
\newcommand{\bigE}{\mathbf{E}}
\newcommand{\bigF}{\mathbf{F}}
\newcommand{\bigGam}{\mathbf{\Gamma}}
\newcommand{\bigXi}{\mathbf{\Xi}}
\newcommand{\bigW}{\mathbf{W}}
\newcommand{\bigM}{\mathbf{M}}

\newcommand{\Pnot}{P_0}
\newcommand{\Rnot}{R_0}
\newcommand{\Pnotp}{P_0^\prime}
\newcommand{\Rnotp}{R_0^\prime}
\newcommand{\ponezero}{p_{10}}
\newcommand{\xonezero}{x_{10}}
\newcommand{\ponezerop}{p_{10}^\prime}
\newcommand{\xonezerop}{x_{10}^\prime}
\newcommand{\ptwozero}{p_{20}}
\newcommand{\xtwozero}{x_{20}}
\newcommand{\ptwozerop}{p_{20}^\prime}
\newcommand{\xtwozerop}{x_{20}^\prime}

\newcommand{\xnot}{\mathbf{x}_0}
\newcommand{\xnotp}{\mathbf{x}_0^\prime}
\newcommand{\dd}[2]{\frac{d #1}{d #2}}

\newcommand{\ddp}[2]{\frac{\partial #1}{\partial #2}}

\newcommand{\tr}{ {\rm Tr} }

\newcommand{\no}{\nonumber}

\newcommand{\ti}{\tilde}

\newcommand{\T}{^{\mathrm{T}}}
\newcommand{\bW}{\mathbf{W}}
\newcommand{\bp}{ {\bf p} }

\newcommand{\bq}{ {\bf q} }

\newcommand{\bx}{ {\bf x} }

\newcommand{\bP}{ {\bf P} }
\newcommand{\bD}{ {\bf D} }
\newcommand{\bR}{ {\bf R}}

\newcommand{\bV}{\mathbf{V}}

\newcommand{\bX}{\mathbf{X}}
\newcommand{\bY}{\mathbf{Y}}
\newcommand{\bC}{\mathbf{C}}
\newcommand{\bS}{\mathbf{S}}
\newcommand{\bLam}{\bm{\Lambda}}

\newcommand{\bM}{\mathbf{M}}
\newcommand{\bJ}{\mathbf{J}}
\newcommand{\bSig}{\bm{\Sigma}}
\newcommand{\bzero}{\mathbf{0}}
\newcommand{\bone}{\mathbf{1}}

\newcommand{\eqr}[1]{Eq.~\eqref{eq:#1}}

\newcommand{\eql}[1]{\label{eq:#1}}

\newcommand{\mL}{\mathcal{L}}

\begin{document}
\preprint{AIP/123-QED}

\title{Nonadiabatic semiclassical dynamics in the mixed quantum-classical initial value representation}

\author{Matthew S. Church}
\affiliation{Department of Chemistry and Chemical Biology, 
Cornell University, Ithaca, New York, 14853, USA}
\author{Timothy J. H. Hele}
\affiliation{Department of Chemistry and Chemical Biology, 
Cornell University, Ithaca, New York, 14853, USA}
\affiliation{Present address: Cavendish Laboratory, JJ Thomson Avenue, Cambridge University, CBS 0HE, UK}
\author{Gregory S. Ezra}
\affiliation{Department of Chemistry and Chemical Biology, 
Cornell University, Ithaca, New York, 14853, USA}
\author{Nandini Ananth}
\email{na346@cornell.edu}
\affiliation{Department of Chemistry and Chemical Biology, 
Cornell University, Ithaca, New York, 14853, USA}

\date{\today}
\begin{abstract} 
We extend the Mixed Quantum-Classical Initial Value 
Representation (MQC-IVR), a semiclassical
method for computing real-time correlation functions,
to electronically nonadiabatic systems using the 
Meyer-Miller-Stock-Thoss (MMST) Hamiltonian to treat 
electronic and nuclear degrees of freedom (dofs) within 
a consistent dynamic framework.
We introduce an efficient symplectic integration 
scheme, the MInt algorithm, for numerical time-evolution 
of the nuclear and electronic phase space variables
as well as the Monodromy matrix, under the non-separable
MMST Hamiltonian. 
We then calculate the probability of transmission through a 
curve-crossing in model two-level systems
and show that in the quantum limit MQC-IVR is in good 
agreement with the exact quantum results, whereas 
in the classical limit the method yields results in keeping
with mean-field approaches like the 
Linearized Semiclassical IVR.
Finally, exploiting the ability of MQC-IVR to quantize different
dofs to different extents, we present a detailed
study of the extents to which quantizing the 
nuclear and electronic dofs improves numerical convergence 
properties without significant loss of accuracy.
\end{abstract}

\maketitle

%\section*{To do}
%\begin{enumerate}
%\item Abstract
%%\item Prose introduction
%%\item Symplecticity algebra
%\item Mention symplecticity issues with other algorithms? $G$ and $\theta$ relation?
%\item Time step for new algorithm vs Adams propagator?
%%\item Evaluate monodromy matrices for Richardson's algorithm?
%\item \timcom{Checking monodromy algebra}
%\item \timcom{Broken/repeated references}
%\end{enumerate}

%%%%%%%%%%%%%%%
%%% INTRODUCTION %%%
%%%%%%%%%%%%%%%
\section{\label{sec:level1}Introduction} 
The development of theoretical methods for
the simulation of electronically nonadiabatic 
processes %electronic and nuclear motion.\cite{tul12a,alt16a} 
remains a central challenge in the effort %our efforts
to understand the mechanisms of 
photochemical reactions,~\cite{dom12a} charge transfer in complex chemical
and biological systems,~\cite{ree09a,gra96a,gra05a,ham10a} and hot-electron generation 
via inelastic scattering.~\cite{gol15a,kru15a}

Over the past two decades, several methods for 
the simulation of nonadiabatic processes have 
been developed including exact quantum 
time-propagation,\cite{top96,bec00a,hel16c} 
the symmetrical quasi-classical windowing method,~
\cite{cot16b} 
mixed quantum-classical Liouville methods~\cite{kap06a,mar98a,mar16a},
 and surface 
hopping.~\cite{tul90a,pre97a,jas02a,her95a,wu07a,lan12a,zim14a,whi15a}
%\nacomment{Matt, add refs for empty cite above, 
%also pull from my paper w/ tom 2010 for better surface
%hopping references - these don't make sense}
In addition, approximate path-integral based methods
such as ring polymer molecular 
dynamics~\cite{cra04a,hab13,hel15a,hel15b,hel17a} 
and centroid molecular dynamics\cite{vot96a} have 
also been extended to nonadiabatic 
systems.~\cite{shu12a,ana13a,
duk15a,duk16a,sha17a,lia02a,mil12a,men11a,ana10a,men14a} 
However, while exact quantum methods are limited to 
a small number of degrees of freedom (dofs), 
the more approximate methods fail to capture 
nuclear quantum coherence effects.

Semiclassical (SC) methods for the calculation of 
real-time correlation functions, like 
the Double Herman-Kluk (DHK) Initial Value 
Representation (IVR) \cite{her84a,mil01a,tho04a,
kay94a,kay05a}, 
accurately describe both electronic and nuclear 
coherence effects in nonadiabatic 
systems~\cite{sun97a,cor00a,ana07a,mil09a,tao13a,ago15a}.
Unfortunately, much like exact quantum methods, 
the high computational cost of numerically converging 
oscillatory integrals has limited these methods to 
low-dimensional systems. Efforts to mitigate the sign
problem have led to the development of more approximate methods
such as the linearized (LSC)-IVR~\cite{wan98a,sun98a,liu15a,shi04a} 
that fail to capture quantum coherence effects,
%\nacomment{check citations, we are missing some of eitan geva's work}
and various forward-backward (FB) methods that are either 
less accurate or computationally expensive.~\cite{mak02a,keg07a,
mak98a,sun99a,mil98a,wan00a,sha99a,gel01a,tho99a}
%\nacomment{split the above citations into references 
%like FBSD that are actually more approximate and 
%FB-IVR/EFB-IVR etc that are equally accurate
%and equally expensive}
%and the inclusion of modified Filinov phase 
%filtratio
 The recently-introduced 
Mixed Quantum-Classical (MQC)-IVR method~\cite{ant15a,chu17a}
employs a modified Filinov filtration (MFF) 
scheme~\cite{cor00a,cor01a,ant15a,chu17a,fil86a,mak87a,mak88a,
spa96a,spa96b,her97a,wal96a,sun98b,tho01a,spa05a} to damp the 
oscillatory phase of the integrand and has been shown to 
improve numerical convergence without significant loss 
of accuracy.~\cite{ant15a,chu17a}
Specifically, the filtering parameters employed in MQC-IVR
modify the extent to which a particular dof 
contributes to the overall phase of the integrand, effectively
controlling the `quantumness' of that mode.~\cite{chu17a}

In this paper we extend MQC-IVR to the simulation 
of nonadiabatic processes using the Meyer-Miller-Stock-Thoss
(MMST)~\cite{mey79a,sto97a} mapping to obtain a continuous 
Cartesian variable representation of both the electronic
and nuclear dofs. We begin by introducing 
an efficient symplectic integration scheme, the MInt algorithm, for 
classical trajectory propagation under the non-separable 
MMST Hamiltonian. We then calculate the transmission probability
using MQC-IVR in a series of model two-level systems with a single
curve crossing. We numerically demonstrate that in the limit of 
a weak filter MQC-IVR agrees well with exact quantum results,
and as the filter strength is increased MQC-IVR results start 
to resemble mean-field methods like the LSC-IVR.
We also undertake a systematic investigation of the balance
between accuracy and efficiency achieved by quantizing the 
nuclear and electronic dofs to different extents.

This paper is organized as follows. In section II we briefly
review the MQC-IVR theory and provide an overview of the 
MInt algorithm. Section III describes the model systems 
studied here and section IV outlines simulation details.
Results are discussed in Section V and we present our 
conclusions in Section VI.

%%%%%%%%%%%
%%% THEORY %%%
%%%%%%%%%%%
\section{\label{sec:level2}Theory}

%%% MQC-IVR THEORY %%%
\subsection{MQC-IVR}
The quantum real-time correlation function\cite{nit06a,tuc10a} 
between two operators $\opA$ and $\opB$ is defined as
\begin{align}\eql{qcorr}
C_{AB}(t)=\Tr\left[\opA e^{\frac{i}{\hbar}\opH t}\opB e^{-\frac{i}{\hbar}\opH t}\right],
\end{align}
where $\opH$ is the system Hamiltonian. 
For the remainder of the paper we use atomic units 
where $\hbar=1$. The MQC-IVR correlation function is 
derived by using the Herman-Kluk (HK-IVR) approximation
for the forward and backward time-evolution operators
in \eqr{qcorr}, followed by a change of variables,
and an MFF of the resulting integrand. 
The final expression is given by~\cite{chu17a}
\begin{align}\eql{DF}
C_{AB}(t)=&\frac{1}{(2\pi)^{2N}}\dx\znot\dx\znotp\;\braket{\znot|\opA|\znotp}\nonumber\\
&\times e^{i\left[S_t(\znot)-S_t(\znotp)\right]}D_t\left(\znot,\znotp;\mathbf{c},\gamz,\gamt\right)\nonumber\\
&\times\braket{\ztp|\opB|\zt}e^{-\frac{1}{2}\mathbf{\Delta}_{\znot}\T\mathbf{c}\mathbf{\Delta}_{\znot}^{}},
\end{align}
where $\znot=(\mathbf{R}_0,\xnot,\mathbf{P}_0,\pnot)$ and 
$\znotp=(\mathbf{R}_0^\prime,\xnotp,\mathbf{P}_0^\prime,\pnotp)$ 
are a pair of initial phase space vectors containing both 
nuclear $(\mathbf{R},\mathbf{P})$ and 
electronic $(\mathbf{x},\mathbf{p})$ variables
associated with classical trajectories of length $t$
and action $S_t(\znot)$ and $S_t(\znotp)$, respectively. 
The full dimensionality of the system is given by $N=F+G$ 
where $F$ and $G$ are the dimensionality of the electronic
and nuclear phase space vectors, respectively.
The phase space displacement between the trajectory 
pair at time zero is given by 
$\mathbf{\Delta}_{\znot}=\znotp-\znot$. 
The functional form of the prefactor, 
$D_t\left(\znot,\znotp;\mathbf{c},\gamz,\gamt\right)$,
is provided in Appendix A. The coherent state 
wavefunctions in momentum and position space are given by 
\begin{align}
\braket{\mathbf{\tilde{P}}\ptil|\zt}=&\left(\frac{1}{\det|\gamt|\pi^N}\right)^\frac{1}{4}e^{-\frac{1}{2}(\mathbf{\tilde{P}}-\mathbf{P}_t)\T
\gamt^{-1}(\mathbf{\tilde{P}}-\mathbf{P}_t)-i\mathbf{\tilde{P}}\T\mathbf{R}_t}\nonumber\\
\times&e^{-\frac{1}{2}(\mathbf{\tilde{p}}-\mathbf{p}_t)\T(\mathbf{\tilde{p}}-\mathbf{p}_t)-i\mathbf{\tilde{p}}\T\mathbf{x}_t}
\eql{mom}
\end{align}
and
\begin{align}
\braket{\mathbf{\tilde{R}}\qtil|\zt}=&\left(\frac{\det|\gamt|}{\pi^N}\right)^\frac{1}{4}e^{-\frac{1}{2}(\mathbf{\tilde{R}}-\mathbf{R}_t)\T
\gamt(\mathbf{\tilde{R}}-\mathbf{R}_t)+i\mathbf{P}_t\T(\mathbf{\tilde{R}}-\mathbf{R}_t)}\nonumber\\
\times&e^{-\frac{1}{2}(\mathbf{\tilde{x}}-\mathbf{x}_t)\T(\mathbf{\tilde{x}}-\mathbf{x}_t)+i\mathbf{p}_t\T(\mathbf{\tilde{x}}-
\mathbf{x}_t)}\eql{poswf},
\end{align}
respectively; and the elements of the $G\times G$ 
diagonal width matrix, $\gamt$, determine the 
spread of the nuclear coherent state in phase 
space at time $t$.

The extent of MFF is controlled by the 
elements of the $2N\times 2N$ diagonal 
matrix of Filinov parameters,
\begin{align}
\mathbf{c}=
\begin{pmatrix}
\cq	&	\mathbb{O}	\\
\mathbb{O}		&	\cpe
\end{pmatrix},
\end{align}
where the subscripts $(\mathbf{q},\bm\rho)$ represent 
the generalized positions and momenta of all $N$ 
dofs, and $\mathbb{O}$ is the null matrix.
The $i^\text{th}$ diagonal element of the
$N\times N$ matrices $\cpe$ and 
$\cq$ regulate momentum and position displacements 
of the $i^\text{th}$ dof 
at time $t=0$.
In the limit $\cpe,\cq\rightarrow 0$, the MQC-IVR 
expression reduces to the standard DHK-IVR 
formulation of the real-time correlation function 
and in the limit $\cpe,\cq\rightarrow \infty$, trajectory 
displacements 
are constrained to $\mathbf{\Delta}_{\znot}=\mathbf{0}$, 
resulting in a classical average,
\begin{align}
C_{AB}(t)=\frac{1}{(2\pi)^N}\dx\znot\braket{\znot|\opA|\znot}\braket{\zt|\opB|\zt},
\end{align}
the Husimi-IVR~\cite{}. 
By choosing intermediate values of the Filinov parameters
for different system modes it is possible to tune the quantumness
of individual modes; an optimal choice can significantly 
accelerate numerical convergence without loss of accuracy.

\subsection{MMST Hamiltonian and the MInt Algorithm}
The MMST Hamiltonian\cite{mey79a,sto97a} for a 
general $F$-level system is given by
\begin{align}\eql{mmst}
H=&\frac{1}{2}\bigP\T\bm{\mu}^{-1}\bigP+\frac{1}{2}\pe\T \bigV(\bigR)\pe\nonumber\\
+&\frac{1}{2}\xe\T\bigV(\bigR)\xe-\frac{1}{2}\Tr\left[\bigV(\bigR)\right],
\end{align}
where  $\bigV(\bigR)$ is the $F\times F$ diabatic 
electronic potential energy matrix and $\bm{\mu}$ 
is the $G\times G$ diagonal matrix of nuclear masses. 
The coupling between nuclear positions and the electronic 
dofs in \eqr{mmst} makes 
it challenging to numerically time-evolve classical 
equations of motion while preserving the 
symplectic property of Hamiltonian systems. 

%Although the original equations of motion for the MM Hamiltonian\cite{mey79b} showed that exact 
%evolution under this Hamiltonian would conserve the total electronic probability (be unitary), numerical studies 
%found population `leakage' and claimed that the dynamics was not unitary.\cite{cor01a,tho99a} 

%\nacomment{zero-point energy leakage is the phenomenon observed
%when {\it individual} electronic state oscillators lose zero-point 
%i.e. the mapping subspace is not properly conserved -- this
%occurs because we are using the MMST as a classical Hamiltonian
%and is independent of the conservation of {\it total} electronic
%probability}
Here we introduce the MInt algorithm for time evolution under
the MMST Hamiltonian in \eqr{mmst} that exactly conserves total electronic 
probability (unitarity) and symplecticity 
%(to within floating point precision) 
independently of time-step size.
We provide a detailed study of this algorithm 
and its properties in Appendix~\ref{ap:alg}.

First we
establish our notation. Hamiltonian evolution is formally\cite{lei04a}
\begin{align}
\dd{}{t} \bz = \bJ \nabla_\bz H(\bz) \eql{hamev}
\end{align}
where $\bJ$ is the structure matrix,
\begin{align}
\bJ = \begin{pmatrix}
\mathbb{O} & \mathbb{I} \\
-\mathbb{I} & \mathbb{O}
\end{pmatrix},
\end{align}
and $\mathbb{I}$ is the identity matrix.
This is equivalent to use of the Poisson 
bracket, $\{\cdot,H(\bz) \}$, since for an arbitrary observable $A$,
\begin{align}
\dd{}{t}A = & (\nabla_{\bz} A) \T \dd{\bz}{t} \no \\
= & (\nabla_{\bz} A)\T \bJ \nabla_\bz H(\bz) \no \\
= & \{A,H(\bz)\}.
\end{align}
In this notation, the Monodromy matrix is given by
\begin{align}
\bM \equiv \dd{\bz_t}{\bz_0} \eql{mondef},
\end{align}
such that the symplecticity criterion is\cite{lei04a}
\begin{align}
\bM\T \bJ^{-1} \bM = \bJ^{-1} \eql{sympcrit}.
\end{align}
We note that this is a stronger condition than 
conservation of volume in phase space 
(Liouville's theorem) which only 
requires $\det| \bM | = 1$.

To construct a symplectic method, we exploit the 
property that exact evolution under a series of 
sub-Hamiltonians gives approximate evolution 
under the total Hamiltonian that is exactly 
symplectic.\cite{lei04a} This scheme is used 
to construct the conventional Velocity Verlet 
algorithm and more complicated algorithms\cite{mil02a} 
such as partitioning the potential energy 
into fast and slowly-varying components.\cite{tuc10a,tuc92a} 
Here, we partition the Hamiltonian in 
\eqr{mmst} into two sub-Hamiltonians,
\begin{subequations}
\begin{align}
H&=H_1+H_2,\\
H_1&=\frac{1}{2}\bigP\T\bm{\mu}^{-1}\bigP, \eql{h1} \\
H_2&=\frac{1}{2}\pe\T \bigV(\bigR)\pe+\frac{1}{2}\xe\T\bigV(\bigR)\xe-\frac{1}{2}\Tr\left[\bigV(\bigR)\right]. \eql{h2}
\end{align}
\end{subequations}
We then define a flow map, $\Phi_{H_i, t}$, corresponding 
to exact evolution [\eqr{hamev}] for 
timestep $t$ under Hamiltonian $H_i$. 
The flow map is simply a function which takes 
as input phase space coordinates $\bz$,
and returns the time-evolved values 
under a specified dynamics. 
In this notation, exact evolution under the MMST 
Hamtiltonian is formally $\bz_t = \Phi_{H,t}(\bz_0)$. 
We define the MInt algorithm as an approximate 
flow map, $\Psi_{H,\Delta t}$, which is a series 
of exact evolutions under the sub-Hamiltonians 
of \eqr{h1} and \eqr{h2},
\begin{align}\eql{flowmap}
\Psi_{H,\Delta t}:=\Phi_{H_1,\Delta t/2}\circ\Phi_{H_2,\Delta t}\circ\Phi_{H_1,\Delta t/2},
\end{align}
where the circles represent the 
composition operation, $f\circ g(\bz):=f(g(\bz))$. 
In words, \eqr{flowmap} describes time evolution
of the system under $H_1$ for half a time step, under
$H_2$ for a full time step, and under $H_1$ again 
for half a time step. As each sub-evolution is 
symplectic, the total evolution will also be 
symplectic.~\cite{lei04a} To confirm this, 
in Appendix~\ref{ap:symp} we prove symplecticity 
directly by evaluating \eqr{sympcrit} for the 
MInt algorithm.

We note that while Liouvillians are commonly used 
to construct symplectic algorithms and to discuss 
time-evolution in general, exact evolution under 
a series of Liouvillians is not necessarily symplectic, 
unless each Liouvillian corresponds to exact evolution 
under a Hamiltonian.~\cite{zwa01a,tuc10a} 
For completeness the MInt algorithm is given in 
the Liouvillian formalism in Appendix~\ref{ap:liou}, 
and compared against a recently-proposed algorithm 
for evolution under the MMST Hamiltonian~\cite{ric16b} 
that is only symplectic in the limit of an infinitely 
small time step.

Evolution under $H_1$ is free particle motion,
\begin{align}\eql{symp1}
\dot{R}_k=\frac{\partial H_1}{\partial P_k}=\frac{P_k}{\mu_{kk}},
\end{align}
for the $k^\text{th}$ nuclear co-ordinates, with all 
other variables fixed. Integrating \eqr{symp1} 
for half
a time step, $\Delta t/2$, yields
\begin{align}\eql{Rsol}
    R_k(\Delta t/2)=R_k(0)+\frac{P_k(0)\Delta t}{2\mu_{kk}}.
\end{align}
%All subsequent steps are assumed to be 
%repeated for all $k$ nuclear components.

For evolution under $H_2$, 
\begin{subequations}
\begin{align}
\dot{\xe}=&\frac{\partial H_2}{\partial\pe}=\bigV(\bigR)\pe \eql{dxdt}, \\
\dot{\pe}=&-\frac{\partial H_2}{\partial\xe}=-\bigV(\bigR)\xe \eql{dpdt}, \\
\dot{P}_k=& -\frac{\partial H_2}{\partial R_k} \no \\
=&-\frac{1}{2}(\xe-i\pe)\T\bigV_k(\bigR)(\xe+i\pe)\nonumber\\
&+\frac{1}{2}\Tr\left[\bigV_k(\bigR)\right], \eql{dpdt2}
\end{align}\eql{h2eq}%
\end{subequations}
with $\bR$ fixed, and we define the gradient 
$\bigV_k(\bigR):=\frac{\partial}{\partial R_k}\bigV(\bigR)$. 
To solve \eqr{h2eq} we note that $\dot \bx$ and $\dot \bp$ 
are not dependent on $ \bP$, but $\dot \bP$ is dependent on 
$ \bx$ and $ \bp$. We can therefore solve for 
$\bx(t)$ and $ \bp(t)$, $0\leq t\leq \Delta t$, and 
substitute this solution into \eqr{dpdt2} 
to find $\bP(\Delta t)$. 

The motion of the electronic positions and momenta is therefore given by\cite{ric16b,hel16c}
\begin{align}
[\xe(\Delta t)+i\pe(\Delta t)]=e^{-i\bigV(\bigR)\Delta t}[\xe(0)+i\pe(0)]. \eql{xpev}
\end{align}
By substituting \eqr{xpev} into \eqr{dpdt2} we obtain 
an expression for nuclear momentum evolution:
\begin{widetext}
\begin{align}
P_k(\Delta t) = P_k(0)-\frac{1}{2} \int_0^{\Delta t} dt \left\{[\xe(0)-i\pe(0)]\T e^{+i\bigV(\bigR) t} \bV_k(\bR) 
e^{-i\bigV(\bigR) t}[\xe(0)+i\pe(0)]- \tr[ \bV_k(\bR)]\right\}. \eql{nucpev}
\end{align}
\end{widetext}
The above equation can be solved analytically, 
as discussed in Appendix~\ref{ap:alg}.
We therefore name the algorithm the 
MInt algorithm as the 
nuclear \textbf{M}omentum \textbf{Int}egral 
over time in \eqr{nucpev} is solved exactly. 
In Appendix~\ref{ap:alg} we also show 
how evolution of the Monodromy
matrix under $\Psi_{H,\Delta t}$ can 
be computed exactly. The evolution will be exactly 
symplectic, satisfying \eqr{sympcrit} for 
any time step (although for very 
large time steps the evolution may 
become a poor approximation to exact 
evolution under $H$).
%In numerical tests we find that the 
%symplecticity criterion is conserved 
%to within machine precision ($\sim 10^{-16}$) 
%for each time step (i.e.\ the element
% of $\bM\T \bJ^{-1} \bM-\bJ^{-1}$ with the greatest
% magnitude at the 
%end of a trajectory is $\sim 10^{-16} N_t$ 
%where $N_t$ is the number of time steps). 

As discussed further in Appendix~C, the MInt 
algorithm is symmetric and time-reversible, 
both properties of exact Hamiltonian evolution. 
Like the Velocity Verlet algorithm, it is 
second order in time step $\Delta t$, and will 
therefore conserve energy with fluctuations of 
$\mathcal{O}(\Delta t^2)$ without drifting. 
The algorithm is also explicit and, being symplectic, 
automatically satisfies Liouville's theorem. 
In addition, as noted for exact evolution under 
the MMST Hamiltonian,\cite{mey79b} the MInt 
algorithm exactly conserves $\mathcal{G}:= 
\bx\T\bx + \bp\T\bp$ and is therefore unitary, 
i.e.\ conserves total electronic 
probability,~\cite{mey79b} 
\begin{align}
\sum_{n=1}^F \mathcal{P}_n 
= \frac{1}{2} \sum_{n=1}^F x_n^2 + p_n^2 - 1,
\end{align}
for any length of time step. It is also invariant to the overall phase (or angle) of the mapping variables, i.e.\ the transformation 
\begin{align}
(\ti \bx + i\ti \bp) = e^{-i\theta} (\bx + i \bp) \eql{theta}
\end{align}
where $\theta$ is a scalar.
We note that this algorithm immediately extends 
to Hamiltonians containing a sum of Meyer-Miller-like 
terms such as the ring polymer Hamiltonians in 
Ref.~\onlinecite{ric16b}.

\section{Model Systems}
\begin{figure}
\includegraphics[scale=0.63]{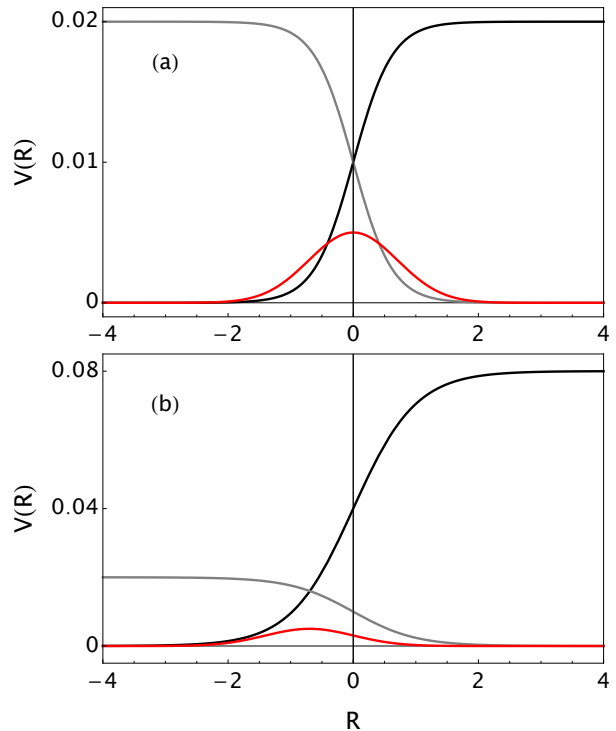}
\caption{Elements of the diabatic electronic potential energy matrix for (a) model 1 and (b) model 2
are plotted as a function of the nuclear position: $V_{11}(R)$ (black), $V_{22}(R)$ (grey) and
$V_{12}(R)$ (red).}
\label{pesurf}
\end{figure}

We test MQC-IVR on previously-used model 2-level systems
with one nuclear dof.~\cite{ana07a}
Model 1 has diabatic electronic potential energy matrix
elements given by
\begin{subequations}
\begin{align}
V_{11}(R)&=V_0\left(1+\tanh\left(\alpha_1 R\right)\right)\\
V_{22}(R)&=V_0\left(1-\tanh\left(\alpha_1 R\right)\right)\\
V_{12}(R)&=a e^{-bR^2},
\end{align}
\end{subequations}
with $V_0=0.01$, $\alpha_1=1.6$, $a=0.005$, and $b=1.0$. 
Model 2 is an asymmetric version of 
model 1,
\begin{subequations}
\begin{align}
V_{11}(R)&=V_1\left(1+\tanh\left(\alpha_2 R\right)\right)\\
V_{22}(R)&=V_0\left(1-\tanh\left(\alpha_2 R\right)\right)\\
V_{12}(R)&=a e^{-b(R+f)^2},
\end{align}
\end{subequations}
with the same parameters as before and $V_1=0.04$, $\alpha_2=1.0$, and $f=0.7$. Plots of the diabats and
couplings for each model are provided in Fig.~\ref{pesurf}.

\section{Simulation Details}
We compute a real-time correlation function 
as defined in \eqr{DF} for a system 
initially in a nuclear coherent state occupying 
electronic state 1. Operator $\opA$ 
is defined as 
\begin{align}
\opA=\ket{\psi_i}\bra{\psi_i}=\ket{P_iR_i1_10_2}\bra{P_iR_i1_10_2},
\end{align}
where $(P_i,R_i)$ denotes the center 
of an initial nuclear coherent state.
The subscripts of $(1_1,0_2)$ label the electronic state 
while a 0 or 1 indicates a ground state or first excited state
configuration in the mapping variables corresponding to that state, respectively.
The corresponding initial position-space wavefunction is 
then given by
\begin{align}
\braket{Rx_1x_2|\psi_i}&=\left(\frac{\gamma}{\pi}\right)^\frac{1}{4}
e^{-\frac{\gamma}{2}(R-R_i)^2+iP_i(R-R_i)}\nonumber\\
&\times\left(\frac{2}{\pi}\right)^\frac{1}{2}x_1 
e^{-\frac{1}{2}(x_1^2+x_2^2)},
\end{align}
with $R_i=-5.0$, and the nuclear coherent state 
width parameter is $\gamma=\gamma_0=\gamma_t=0.25$. 
Simulations are performed with either large incident
kinetic energy, $0.1$, corresponding to initial 
nuclear momentum $P_i=19.9$, or low incident
kinetic energy, $0.03$, where $P_i=10.9$.
The nuclear mass is $1980$.

To compute the particle's distribution of 
final translational momentum at long times, $P_f$,
we define $\opB=\delta(P_f-\hat{P})$.
The MQC-IVR expression for this choice of operators
is
\begin{align}\eql{cpf}
C(P_f)=\lim_{t\rightarrow\infty}&\frac{1}{(2\pi)^6}\dx\znot\dx\znotp\braket{\znot|\psi_i}\braket{\psi_i|\znotp}\nonumber\\
&\times e^{i\left[S_t(\znot)-S_t(\znotp)\right]}D_t\left(\znot,\znotp;\mathbf{c},\gamz,\gamt\right)\nonumber\\
&\times \braket{\ztp|\delta(P_f-\hat{P})|\zt}e^{-\frac{1}{2}\mathbf{\Delta}_{\znot}\T\mathbf{c}\mathbf{\Delta}_{\znot}^{}}.
\end{align}

For model 1, we sample the initial nuclear coordinates with the following correlated sampling distribution,\cite{pan13a}
\begin{align}
\omega_N(\Pnot,\Rnot,\Pnotp,\Rnotp)=&|\braket{\bar{P}_0\bar{R}_0|P_iR_i}|^2\nonumber\\
\times&e^{-\frac{c_P}{2}\DPnot^2}e^{-\frac{c_R}{2}\DRnot^2},
\end{align}
where the bars represent mean variables [e.g. $\bar{P}_0=\frac{1}{2}(\Pnotp+\Pnot)$]. The initial 
coordinates of oscillator 1 are sampled from
\begin{align}
\omega_1(\ponezero,\xonezero,\ponezerop,\xonezerop)=&|\braket{\ponezero\xonezero|1_1}|^2
|\braket{\ponezerop\xonezerop|1_1}|^2\nonumber\\
\times&e^{-\frac{c_{\ponezero}}{2}\Delta_{\ponezero}^2-\frac{c_{\xonezero}}{2}\Delta_{\xonezero}^2},
\end{align}
where the first subscript of the mapping variables indicates the electronic state and the second
subscript indicates the time.
The initial coordinates of oscillator 2 are sampled from
\begin{align}
\omega_2(\ptwozero,\xtwozero,\ptwozerop,\xtwozerop)=&|\braket{\ptwozero\xtwozero|0_2}|^2
|\braket{\ptwozerop\xtwozerop|0_2}|^2\nonumber\\
\times&e^{-\frac{c_{\ptwozero}}{2}\Delta_{\ptwozero}^2-\frac{c_{\xtwozero}}{2}\Delta_{\xtwozero}^2}.
\end{align}
For model 2, we use a different sampling scheme that proves
more efficient,
\begin{align}
\omega(\znot,\znotp)=\left|\braket{\znot|\psi_i}\braket{\psi_i|\znotp}\right|e^{-\frac{1}{2}\mathbf{\Delta}_{\znot}\T\mathbf{c}\mathbf{\Delta}_{\znot}^{}}.
\end{align}

The overlap of the coherent states with operator $\opB=\delta(P_f-\hat{P})$ can be found by inserting a 
momentum identity and using \eqr{mom},
\begin{align}
&\braket{\ztp|\delta(P_f-\hat{P})|\zt}=\left(\frac{1}{\gamma\pi}\right)^\frac{1}{2}e^{-\frac{\gamma}{2}
(P_f-P_t^\prime)^2}\nonumber\\
&\times e^{-\frac{\gamma}{2}(P_f-P_t)^2}e^{i P_f(R_t^\prime-R_t)}\\
&\times\prod_{j=1}^2e^{-\frac{1}{4}(x_{jt}^\prime-x_{jt})^2-\frac{1}{4}(p_{jt}^\prime-p_{jt})^2}
e^{\frac{i}{2}(p_{jt}^\prime+p_{jt})(x_{jt}^\prime-x_{jt})}.\nonumber
\end{align}

For both models we use a time step of 
$\Delta t = 1.5$ a.u. and monitor energy conservation with
a tolerance parameter, $\epsilon=10^{-4}$, such that 
\begin{align}
\left | 1-E(0)/E(t)\right | < \epsilon.
\end{align}
With the MInt algorithm, we find that 
only $\sim 0.1\%$ of trajectories violate
this tolerance in the model systems presented here
and with the time step mentioned above. We use a total simulation
time of 3000 a.u. for the high energy simulations and 4000 a.u.
for the low energy simulations. 
We also track the phase of the prefactor in 
order to select the correct branch of the 
complex square root.
Exact quantum results are obtained by diagonalizing the 
quantum mechanical Hamiltonian
in the Discrete Variable Representation, followed by time-evolution
with a Chebyshev propagation algorithm.\cite{tan07a,ana07a}

For all results presented below, we set the 
position and momentum filtering parameters for a given 
dof to be equal: $\cq=\cpe$. Further,
we take all electronic filtering parameters 
to be equal, thus treating the two electronic
states at the same level of quantization.
For clarity, in the rest of this paper, we 
use $c_\text{nuc}$ and $c_\text{el}$ to indicate the values 
used to filter the nuclear and electronic dofs, respectively.

%%%%%%%%%%%%
%%% RESULTS %%%
%%%%%%%%%%%%
\section{Results}
\begin{figure}
\includegraphics[scale=0.55]{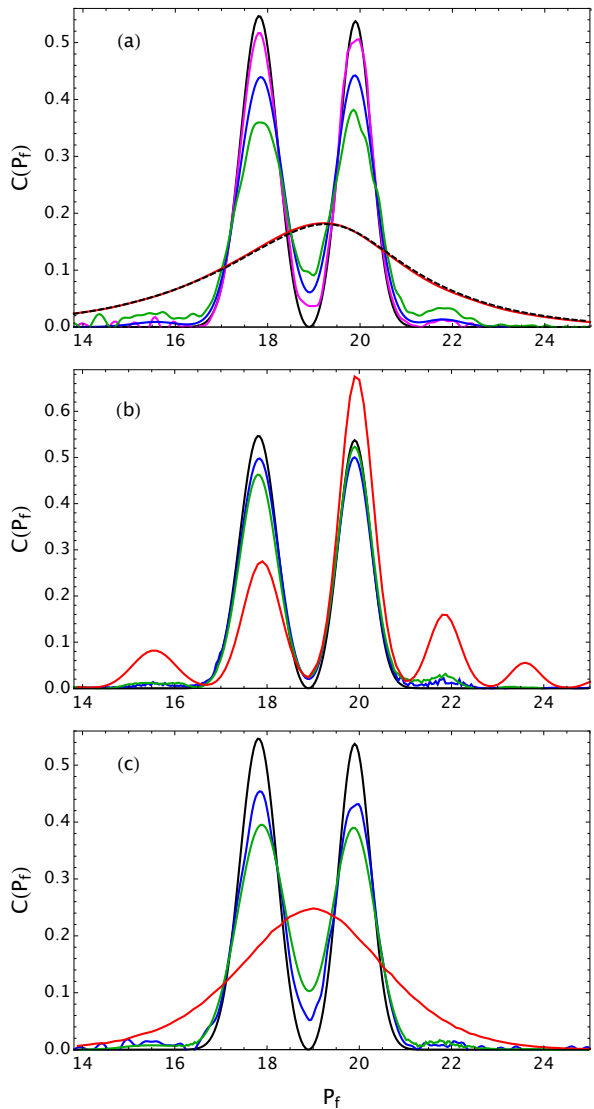}
\caption{The distribution of final nuclear momentum
with model 1 and an incident energy of $0.1\;E_h$. The exact quantum result (black, solid) is shown in each panel
along with 
(a) the Husimi-IVR (black, dashed)
and MQC-IVR where each dof is treated with the same filtering strength: $c=0.01$ (pink), 
$c=0.05$ (blue), $c=0.1$ (green), and $c=10.0$ (red);
(b) the MQC-IVR results where the nuclear filtering parameters are fixed near the quantum limit, 
$c_\text{nuc}=0.01$, and the
electronic filtering parameters are varied: $c_{\text{el}}=0.05$ (blue), $c_\text{el}=0.1$ (green), 
and $c_\text{el}=10.0$ (red);
 (c) MQC-IVR results where the electronic filtering parameters
are fixed near the quantum limit, $c_\text{el}=0.01$, and the nuclear filtering parameters are varied:
$c_\text{nuc}=0.05$ (blue), $c_\text{nuc}=0.1$ (green), and $c_\text{nuc}=10.0$ (red).}
\label{mod1Ehigh}
\end{figure}

\begin{figure}
\includegraphics[scale=0.55]{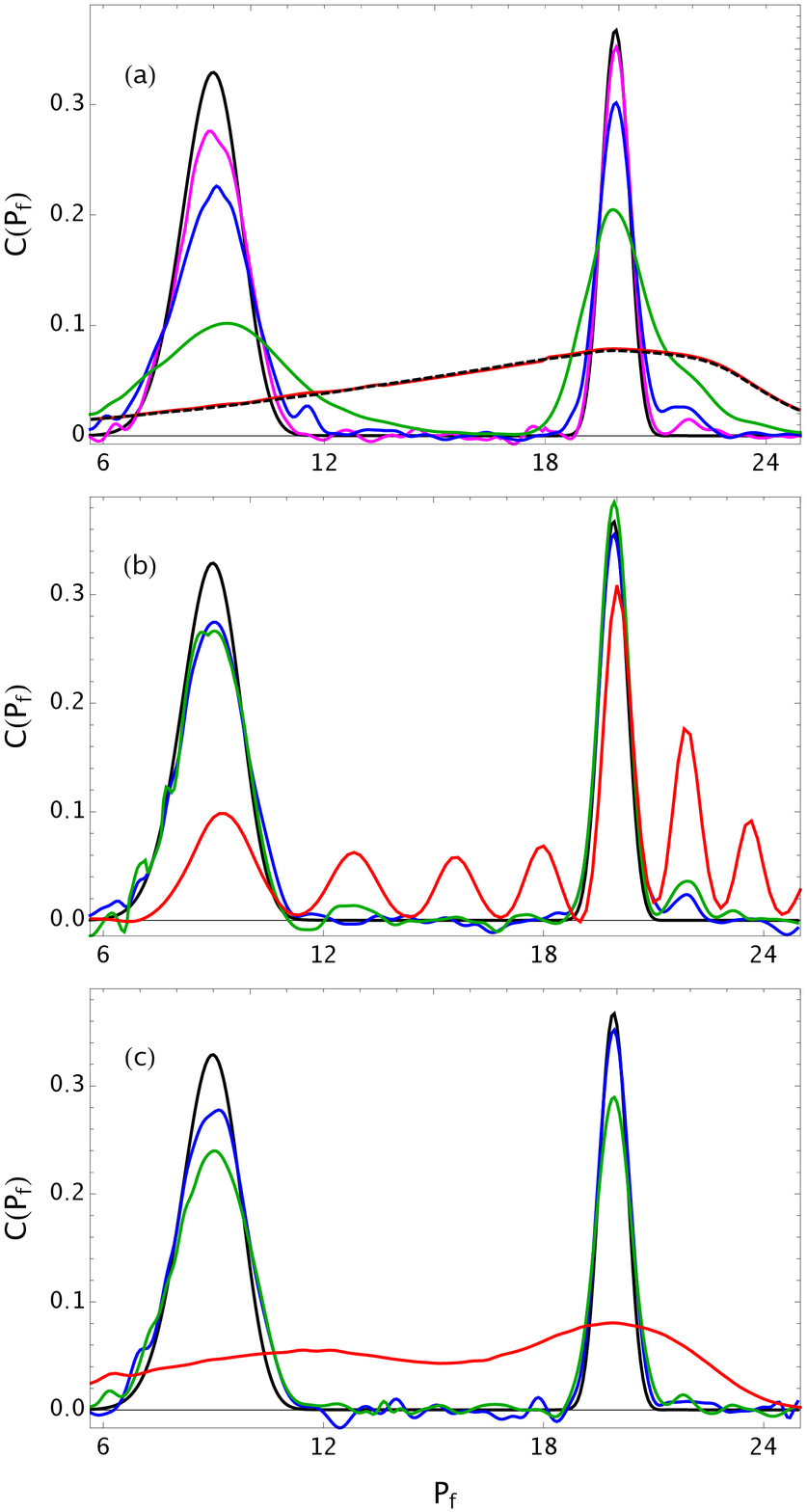}
\caption{The distribution of final nuclear momentum
with model 2 and incident energy of $0.1\;E_h$. The exact quantum result (black, solid) is shown in each panel
along with 
(a) the Husimi-IVR (black, dashed)
and MQC-IVR where each dof is treated with the same filtering strength: $c=0.01$ (pink),
 $c=0.05$ (blue), $c=0.1$ (green), and $c=10.0$ (red);
(b) the MQC-IVR results where the nuclear filtering parameters are fixed near the quantum limit, 
$c_\text{nuc}=0.01$, and the
electronic filtering parameters are varied from $c_\text{el}=0.05$ (blue) to $c_{\text{el}}=0.1$ (green)
and $c_\text{el}=10.0$ (red);
 (c) MQC-IVR results where the electronic filtering parameters
are fixed in the quantum-limit, $c_\text{el}=0.01$, and the nuclear filtering parameters are varied from
$c_\text{nuc}=0.05$ (blue) to $c_\text{nuc}=0.1$ (green) and $c_\text{nuc}=10.0$ (red).}
\label{mod2Ehigh}
\end{figure}

\begin{figure}
\includegraphics[scale=0.55]{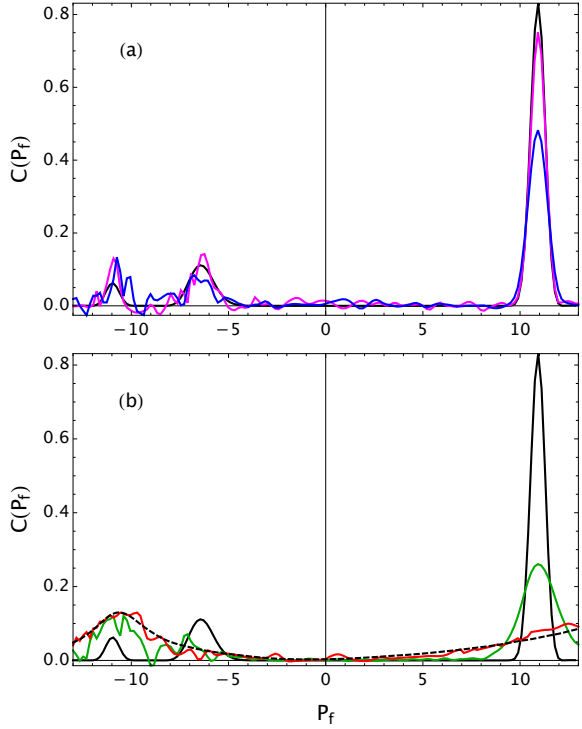}
\caption{The final distribution of nuclear momentum with model 2 and an incident energy of $0.03\,E_h$. In both
panels the
exact quantum result is shown in black along with MQC-IVR results in which each dof is filtered
equally: (a) $c=0.01$ (pink) and $c=0.1$ (blue); (b) $c=1.0$ (green) and $c=10.0$ (red).}
\label{mod2Elow1}
\end{figure}

\begin{figure}
\includegraphics[scale=0.55]{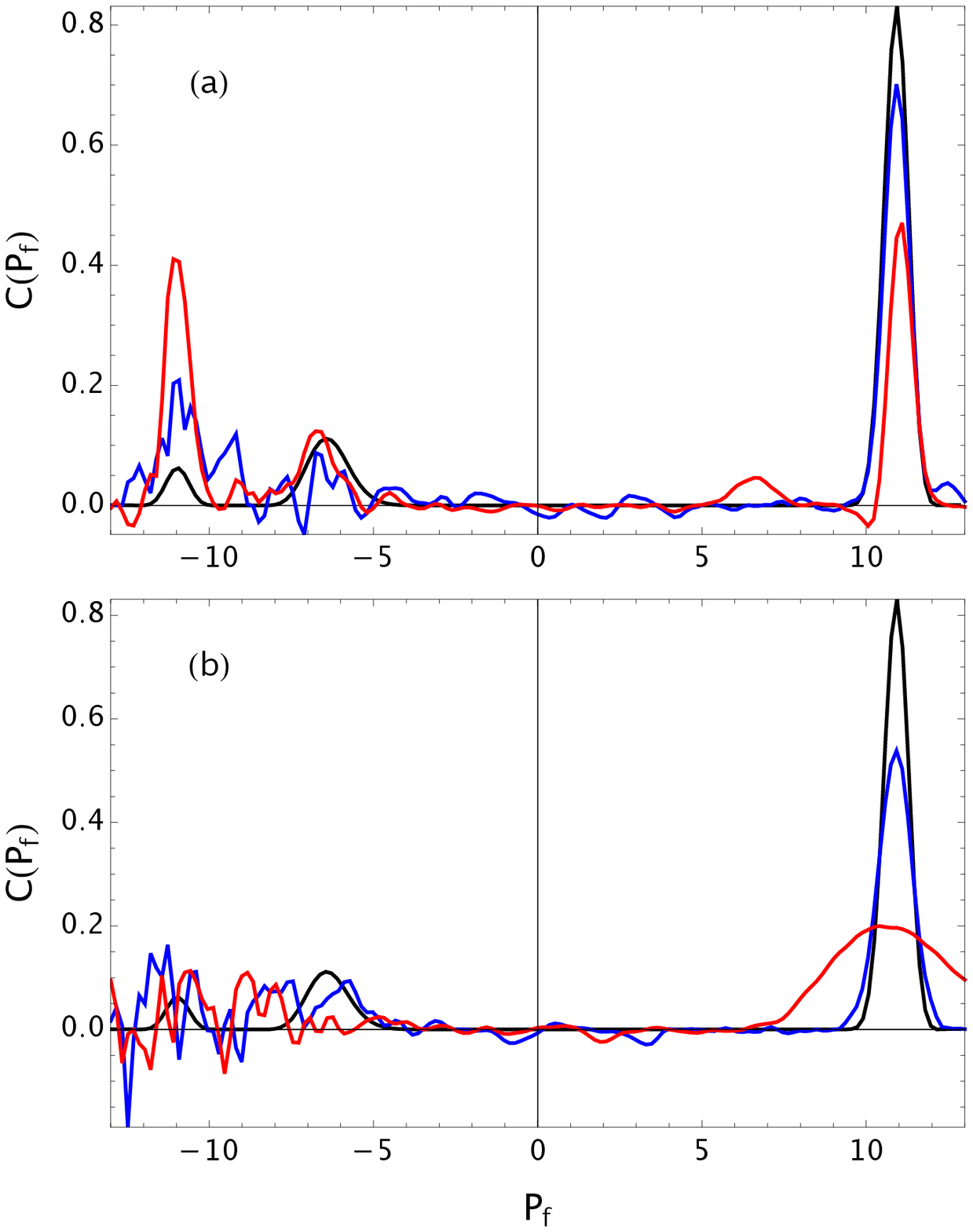}
\caption{The final distribution of nuclear momentum with model 2 and an incident energy of $0.03 \ E_h$. The
exact quantum result is shown in black along with MQC-IVR results where (a) the nuclear filtering parameters are 
fixed near the quantum limit, $c_\text{nuc}=0.01$, and the electronic dofs are treated with
$c_\text{el}=1.0$ (blue) and $c_\text{el}=10.0$ (red); (b) the electronic filtering parameters are fixed near the
quantum limit, $c_\text{el}=0.01$, and the nuclear dofs are treated with $c_\text{nuc}=1.0$ (blue)
and $c_\text{nuc}=10.0$ (red).}
\label{mod2Elow2}
\end{figure}

Here we show the results of using \eqr{cpf} to compute the particle's 
distribution of final nuclear momentum after transmission through
the curve crossing in models 1 and 2. MQC-IVR results obtained for 
model 1 and a high incident energy of $0.1E_h$ are shown in Fig.~\ref{mod1Ehigh}, 
for model 2 with the same high incident energy of $0.1E_h$ are shown in Fig.~\ref{mod2Ehigh},
and for model 2 with a low incident energy of $0.03E_h$ in Fig.~\ref{mod2Elow1} and
Fig.~\ref{mod2Elow2} respectively. All panels show the exact quantum result as a solid black curve.

In Fig.~\ref{mod1Ehigh}(a), Fig.~\ref{mod2Ehigh}(a), Fig.~\ref{mod2Elow1}(a) 
and Fig.~\ref{mod2Elow1}(b), all dofs are equally quantized with 
$c=c_\text{nuc}=c_\text{el}$. As expected, the quantum limit filtering strength
($c=0.01$ shown in pink in the first three figures mentioned) agrees well with 
the transmission peaks of the exact quantum results, 
with slight reduction in peak amplitudes and slight broadening of peak widths. 
The reflection peaks at $P_f=-6.5$ and $P_f=-11.0$ of Fig.~\ref{mod2Elow1}(a) 
in this limit, though noisier than the high-intensity transmission peaks,
also agree well with the exact quantum result, but with a slight over-estimation 
of each signal. Increasing the strength of the filter
(larger values of $c=0.05$ and $c=1.0$ shown in blue and green respectively) 
in each model further 
broadens peak widths and reduces peak amplitudes,
but the discrete quantum peak structure is retained in each case and significantly 
fewer trajectories are required for convergence, as
reported in Tables~\ref{tabM1}-\ref{tabM2LOW}. The 
deviation from exact quantum increases 
as we further increase filtering strength and, as expected, the 
MQC-IVR result collapses to the Husimi-IVR result
[shown in black, dashed in Fig.~\ref{mod1Ehigh}(a), Fig.~\ref{mod2Ehigh}(a)
and Fig.~\ref{mod2Elow1}(b)] when the filter 
strength is $c\geq 10$ [shown in red in Fig.~\ref{mod1Ehigh}(a), Fig.~\ref{mod2Ehigh}(a)
and Fig.~\ref{mod2Elow1}(b)].
\begin{table}
\centering
\begin{tabular}{c c c c c c c}\hline\hline
$c_\text{nuc}$	&	$\;\;\;\;\;$		&	$c_\text{el}$	&	$\;\;\;\;\;$			&	$N_\text{traj}$	&	$\;\;\;\;\;$	&	$ \text{max}\left[\varepsilon(P_f) \right]$\\\hline\hline
$0.01$		&	$\;\;\;\;\;$		&	$0.01$		&	$\;\;\;\;\;$			&$3.2\times 10^9$	&	$\;\;\;\;\;$		&	$3.7\times 10^{-2}$	\\\hline
$0.05$		&	$\;\;\;\;\;$		&	$0.05$		&	$\;\;\;\;\;$			&$5.8\times 10^8$	&	$\;\;\;\;\;$		&	$1.1\times 10^{-1}$	\\\hline
$0.1$		&	$\;\;\;\;\;$		&	$0.1$		&	$\;\;\;\;\;$			&$4.8\times 10^8$	&	$\;\;\;\;\;$		&	$1.9\times 10^{-1}$ \\\hline
$10.0$		&	$\;\;\;\;\;$		&	$10.0$		&	$\;\;\;\;\;$			&$1.5\times 10^6$	&	$\;\;\;\;\;$		&	$4.1\times 10^{-1}$ \\\hline
$0.01$		&	$\;\;\;\;\;$		&	$0.05$		&	$\;\;\;\;\;$			&$7.4\times 10^8$	&	$\;\;\;\;\;$		&	$5.1\times 10^{-2}$ \\\hline
$0.01$		&	$\;\;\;\;\;$		&	$0.1$		&	$\;\;\;\;\;$			&$6.3\times 10^8$	&	$\;\;\;\;\;$		&	$8.4\times 10^{-2}$ \\\hline
$0.01$		&	$\;\;\;\;\;$		&	$10.0$		&	$\;\;\;\;\;$			&$2.4\times 10^7$	&	$\;\;\;\;\;$		&	$2.8\times 10^{-1}$ \\\hline
$0.05$		&	$\;\;\;\;\;$		&	$0.01$		&	$\;\;\;\;\;$			&$1.5\times 10^9$	&	$\;\;\;\;\;$		&	$1.1\times 10^{-1}$	\\\hline
$0.1$		&	$\;\;\;\;\;$		&	$0.01$		&	$\;\;\;\;\;$			&$8.8\times 10^8$	&	$\;\;\;\;\;$		&	$1.6\times 10^{-1}$ \\\hline
$10.0$		&	$\;\;\;\;\;$		&	$0.01$		&	$\;\;\;\;\;$			&$4.8\times 10^8$	&	$\;\;\;\;\;$		&	$3.7\times 10^{-1}$ \\\hline
\end{tabular}
\caption{The number of trajectories required for graphical convergence, $N_\text{traj}$, 
of each MQC-IVR result in Fig.~\ref{mod1Ehigh}. Also listed is the absolute error relative to the
exact quantum result, as averaged over $P_f$.}
\label{tabM1}
\end{table}
\begin{table}
\centering
\begin{tabular}{c c c c c c c}\hline\hline
$c_\text{nuc}$	&	$\;\;\;\;\;$	&	$c_\text{el}$		&	$\;\;\;\;\;$	&	$N_\text{traj}$		&	$\;\;\;\;\;$	&	$ \text{max}\left[\varepsilon(P_f) \right]$\\\hline\hline
$0.01$		&	$\;\;\;\;\;$	&	$0.01$			&	$\;\;\;\;\;$	&	$1.6\times 10^9$	&	$\;\;\;\;\;$	&	$5.7\times 10^{-2}$	\\\hline
$0.05$		&	$\;\;\;\;\;$	&	$0.05$			&	$\;\;\;\;\;$	&	$4.8\times 10^8$	&	$\;\;\;\;\;$	&	$1.1\times 10^{-1}$ \\\hline
$0.1$		&	$\;\;\;\;\;$	&	$0.1$			&	$\;\;\;\;\;$	&	$2.8\times 10^8$	&	$\;\;\;\;\;$	&	$2.3\times 10^{-1}$ \\\hline
$10.0$		&	$\;\;\;\;\;$	&	$10.0$			&	$\;\;\;\;\;$	&	$3.6\times 10^6$	&	$\;\;\;\;\;$	&	$3.1\times 10^{-1}$ \\\hline
$0.01$		&	$\;\;\;\;\;$	&	$0.05$			&	$\;\;\;\;\;$	&	$7.2\times 10^8$	&	$\;\;\;\;\;$	&	$5.5\times 10^{-2}$ \\\hline
$0.01$		&	$\;\;\;\;\;$	&	$0.1$			&	$\;\;\;\;\;$	&	$6.0\times 10^8$	&	$\;\;\;\;\;$	&	$6.3\times 10^{-2}$ \\\hline
$0.01$		&	$\;\;\;\;\;$	&	$10.0$			&	$\;\;\;\;\;$	&	$1.2\times 10^8$	&	$\;\;\;\;\;$	&	$2.4\times 10^{-1}$ \\\hline
$0.05$		&	$\;\;\;\;\;$	&	$0.01$			&	$\;\;\;\;\;$	&	$9.4\times 10^8$	&	$\;\;\;\;\;$	&	$5.5\times 10^{-2}$ \\\hline
$0.1$		&	$\;\;\;\;\;$	&	$0.01$			&	$\;\;\;\;\;$	&	$7.2\times 10^8$	&	$\;\;\;\;\;$	&	$9.0\times 10^{-2}$ \\\hline
$10.0$		&	$\;\;\;\;\;$	&	$0.01$			&	$\;\;\;\;\;$	&	$4.1\times 10^8$	&	$\;\;\;\;\;$	&	$2.9\times 10^{-1}$ \\\hline
\end{tabular}
\caption{The number of trajectories required for graphical convergence, $N_\text{traj}$, 
of each result in Fig.~\ref{mod2Ehigh}. Also listed is the absolute error relative to the
exact quantum result, as averaged over $P_f$.}
\label{tabM2}
\end{table}
\begin{table}
\centering
\begin{tabular}{c c c c c c c}\hline\hline
$c_\text{nuc}$	&	$\;\;\;\;\;$	&	$c_\text{el}$		&	$\;\;\;\;\;$	&	$N_\text{traj}$		&	$\;\;\;\;\;$	&	$\text{max}\left[\varepsilon(P_f) \right]$\\\hline\hline
$0.01$		&	$\;\;\;\;\;$	&	$0.01$			&	$\;\;\;\;\;$	&	$3.0\times 10^9$	&	$\;\;\;\;\;$	&	$8.0\times 10^{-2}$	\\\hline
$0.1$		&	$\;\;\;\;\;$	&	$0.1$			&	$\;\;\;\;\;$	&	$1.5\times 10^9$	&	$\;\;\;\;\;$	&	$3.5\times 10^{-1}$ \\\hline
$1.0$		&	$\;\;\;\;\;$	&	$1.0$			&	$\;\;\;\;\;$	&	$2.6\times 10^8$	&	$\;\;\;\;\;$	&	$5.7\times 10^{-1}$ \\\hline
$10.0$		&	$\;\;\;\;\;$	&	$10.0$			&	$\;\;\;\;\;$	&	$2.2\times 10^6$	&	$\;\;\;\;\;$	&	$7.5\times 10^{-1}$ \\\hline
$0.01$		&	$\;\;\;\;\;$	&	$0.1$			&	$\;\;\;\;\;$	&	$1.7\times 10^9$	&	$\;\;\;\;\;$	&	$1.5\times 10^{-1}$ \\\hline
$0.01$		&	$\;\;\;\;\;$	&	$10.0$			&	$\;\;\;\;\;$	&	$4.5\times 10^7$	&	$\;\;\;\;\;$	&	$4.2\times 10^{-1}$ \\\hline
$0.1$		&	$\;\;\;\;\;$	&	$0.01$			&	$\;\;\;\;\;$	&	$2.4\times 10^9$	&	$\;\;\;\;\;$	&	$2.9\times 10^{-1}$ \\\hline
$10.0$		&	$\;\;\;\;\;$	&	$0.01$			&	$\;\;\;\;\;$	&	$4.5\times 10^8$	&	$\;\;\;\;\;$	&	$6.3\times 10^{-1}$ \\\hline
\end{tabular}
\caption{The number of trajectories required for graphical convergence, $N_\text{traj}$, 
of each result in Fig.~\ref{mod2Elow1} and Fig.~\ref{mod2Elow2}. Also listed is the absolute error relative to the
exact quantum result, as averaged over $P_f$.}
\label{tabM2LOW}
\end{table}

We then present MQC-IVR results where the 
nuclear and electronic dofs are quantized to different extents
by varying $c_\text{el}$ and $c_\text{nuc}$ independently.
In Fig.~\ref{mod1Ehigh}(b), Fig.~\ref{mod2Ehigh}(b), and Fig.~\ref{mod2Elow2}(a) we fix
the nuclear dof in the quantum limit ($c_\text{nuc}=0.01$) and vary the tuning strength
associated with the electronic dofs between $c_\text{el}=0.05$ and $c_\text{el}=10.0$. 
Although the quantum double peak structure is visible in
all cases considered here, as we move towards the classical 
limit ($c_\text{el}=10.0$ shown in red in each case) 
spurious peaks appear and relative
peak intensities change dramatically.
We note that, unlike in Fig.~\ref{mod1Ehigh}(a), Fig.~\ref{mod2Ehigh}(a),
and Fig.~\ref{mod2Elow1}(b), where the peaks merge to the mean-field 
Husimi-IVR result in the classical-limit,
the discrete peak structure is still visible
when only the electronic dofs are treated 
in the classical limit.

Next, in Fig.~\ref{mod1Ehigh}(c), Fig.~\ref{mod2Ehigh}(c), and 
Fig.~\ref{mod2Elow2}(b) we treat the electronic dofs 
in the quantum limit ($c_{\text{el}}=0.01$)
and vary the extent of nuclear quantization 
from $c_\text{nuc}=0.01$ to $c_\text{nuc}=10.0$. We find these 
results are very similar to those 
in Fig.~\ref{mod1Ehigh}(a), Fig.~\ref{mod2Ehigh}(a), Fig.~\ref{mod2Elow1}(a),
and Fig.~\ref{mod2Elow1}(b) where both electronic
and nuclear dofs are equally quantized \textemdash
the spurious peaks that appear in the cases where the electron dofs
are treated in the classical-limit do not appear, instead 
the peaks start to merge with larger $c_\text{nuc}$. This gives 
rise to mean-field like behavior where transmission probability
is highest on an unphysical, average electronic surface.

As mentioned above, Tables~\ref{tabM1}-\ref{tabM2LOW}
report the total number of trajectories required for graphical convergence of each 
MQC-IVR result. Also reported in each table is the maximum absolute error
\begin{align}
\varepsilon(P_f)=\left|C_\text{MQC}(P_f)-C_\text{QM}(P_f)\right|
\end{align}
of each result across all values of $P_f$ along with 
the reduction in computational cost observed for even small values of the 
filtering parameters. This allows us to clearly identify parameter regimes
where the filtering results in improved convergence but little 
reduction in accuracy. For the high energy simulations with models 1 
and 2, an optimal choice of parameters may be 
$c_\text{nuc}=0.01$ and $c_\text{el}=0.05$ or $c_\text{el}=0.1$ where the number of 
trajectories required for convergence is on the order 
of $10^8$ with maximum absolute error on the order of $10^{-2}$. 
More trajectories are required in this parameter regime for the low 
energy simulation of model 2, due to the slower convergence of the 
reflection peaks, but the number of trajectories required
is nearly half that of the weakest filter ($c=0.01$), and 
the maximum absolute error only increases
from $0.08$ to $0.15$. We hypothesize 
that since we are calculating a nuclear observable here,
it is necessary to quantize the nuclear dof to a greater
extent than the electronic dofs. 
This idea is further validated by an observation made 
in the original MQC-IVR implementation~\cite{ant15a} for a
model 2D adiabatic system of coupled oscillators. Specifically,
it was shown that when observing the position of the heavy (more 
classical mode) it was sufficient to quantize just that mode 
and the accuracy of the resulting correlation function was largely
independent of the extent of quantization used to describe
the lighter, unobserved mode.~\cite{ant15a}

\begin{figure}
\includegraphics[scale=0.55]{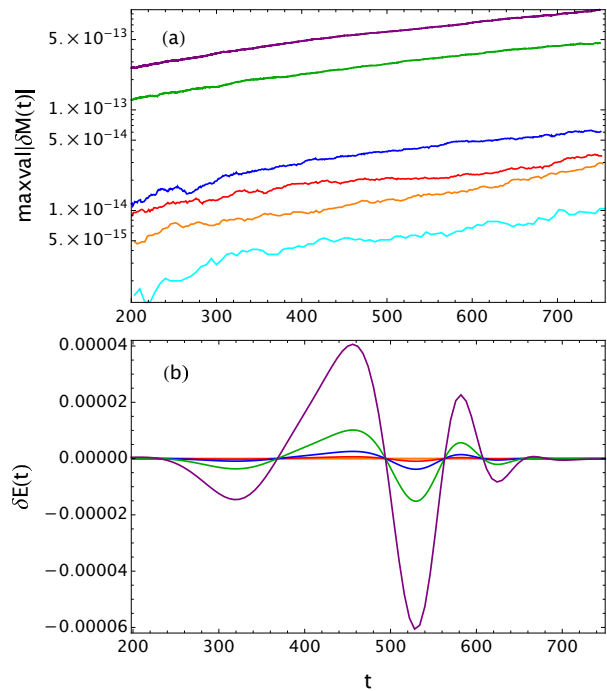}
\caption{A numerical analysis of the MInt algorithm with a single low-energy trajectory for model 2: 
(a) a histogram of the largest element 
of $\delta\mathbf{M}(t)$ as a function of time, and (b) a histogram showing energy conservation as a function of time.
Each color represents a different time step used: $\Delta t=0.05$ (cyan), $0.10$ (orange), $0.75$ (red), $1.5$ (blue),
$3.0$ (green), $6.0$ (purple).}
\label{MIntfig}
\end{figure}

Finally, in Fig.~\ref{MIntfig} we provide numerical evidence 
of two important features of the MInt algorithm: symplecticity
and energy conservation. We monitor symplecticity by tracking the
element of the matrix
\begin{align}
\delta \mathbf{M}(t)=\mathbf{M}_{qq}\T\mathbf{M}_{\rho\rho}^{}-\mathbf{M}_{\rho q}\T\mathbf{M}_{q\rho}^{}-\mathbb{I}
\eql{sympcons}
\end{align}
with the greatest magnitude: a condition derived from \eqr{sympcrit}. 
Our energy conservation criterion is
\begin{align}
\delta E(t) = \left|1-E(t)/E(0)\right|.
\eql{econs}
\end{align}
For a single low-energy trajectory of model 2 we plot the quantity in \eqr{sympcons} in Fig.~\ref{MIntfig}(a)
with a log-scaled y-axis, and the function in
\eqr{econs} in Fig.~\ref{MIntfig}(b). Each quantity is plotted as a function of time as the particle traverses the interaction region, and 
each colored curve represents a different choice of time step ranging from $\Delta t =0.05$ to
$\Delta t=6.0$. The linear growth of the largest element of $\delta \mathbf{M}(t)$ in Fig.~\ref{MIntfig}(a) indicates
that the MInt algorithm is symplectic for both fine and coarse time steps. The fluctuations in $\delta E(t)$ in
Fig.~\ref{MIntfig}(b) oscillate around the true value, and the amplitude is damped with finer time steps: a characteristic
of a symplectic algorithm.

%The fluctuations
%in $\delta E(t)$ in Fig.~\ref{MIntfig}(b) occur when the particle reaches the interaction region near the curve crossing.
%The oscillations are short-lived due to the flatness of the potential energy landscape, but they clearly oscillate about
%the true value, and no energy drift is observed.

\section{Conclusions}
In this article we have successfully extended MQC-IVR to the description of nuclear 
coherence effects in nonadiabatic systems. We have analyzed the effects of treating both electronic
and nuclear dofs under identical and different filtering strengths, and found that there
are parameter regimes in both cases which not only reduce computational expense (as opposed to a minimally
filtered SC correlation function) but also maintain a
qualitative description of the transmission through a curve crossing.

We have also proposed the MInt algorithm for exactly symplectic evolution under the MMST Hamiltonian, 
and discussed its other properties.

In future work we plan to extend nonadiabatic MQC-IVR to multidimensional nonadiabatic systems such as the 
NO scattering problem,\cite{gol15a,kru15a} as well as implement the MInt (or similar) algorithms in other nonadiabatic 
dynamics methods based upon the MMST Hamiltonian.

\section*{Acknowledgements}
The authors acknowledge several helpful discussions with Sergey V. Antipov 
and his contribution to some of the original code used here.
This work was funded, in part, by Army Research Office Grant No. W911NFD-13-1-0102
and an NSF EAGER award No. CHE-1546607. 
In addition, NA acknowledges funding from the Research Corporation for Science Advancement
through a Cottrell Scholar Award, a Sloan Foundation Fellowship, and start-up funding
from Cornell University.
%MSC and NA acknowledge several helpful discussions with Sergey V. Antipov. 
TJHH acknowledges a Research Fellowship from Jesus College, Cambridge.

\appendix
\section{MQC-IVR Prefactor}
The functional form of the prefactor is given by
\begin{align}
&D_t\left(\znot,\znotp;\mathbf{c},\gamz,\gamt\right)\;=\;\det(\frac{1}{2}\gamt^{-1}\G)^\frac{1}{2}\nonumber\\
&\times\det\bigg[\frac{1}{2}(\mppf-i\gamt\mqpf)(\G^{-1}+\mathbb{I})(\mppb\gamt+i\mpqb)\nonumber\\
&+(\gamt\mqqf+i\mpqf)(\frac{1}{2}\gamz^{-1}+\cpe)\G^{-1}(\mppb\gamt+i\mpqb)\nonumber\\
&+\frac{1}{2}(\gamt\mqqf+i\mpqf)(\G^{-1}+\mathbb{I})(\mqqb-i\mqpb\gamt)\nonumber\\
&+(\mppf-i\gamt\mqpf)(\frac{1}{2}\gamz+\cq)\G^{-1}(\mqqb-i\mqpb\gamt)\bigg]^\frac{1}{2}\nonumber,
\end{align}
with diagonal matrix $\G=(\cq+\gamz)\cpe+\cq(\gamz^{-1}+\cpe)$. We define elements of the un-primed trajectory's
 monodromy matrix as $\mathbf{M}_{\alpha\beta}^f=\frac{\partial\alpha_t}{\partial\beta_0}$ and the primed trajectory's
 backward monodromy matrix as $\mathbf{M}_{\alpha\beta}^b=\frac{\partial \alpha_0^\prime}{\partial\beta_t^\prime}$. Note that the
backward monodromy matrix is related to its forward counterpart with the following
identity,
\begin{align}
\mathbf{M}^{b}\;=\;(\mathbf{M}^{f\prime})^{-1}\;=\;
\begin{pmatrix}
\mathbf{M}_{\rho\rho}^{f\T\prime}	&	-\mathbf{M}_{q\rho}^{f\T\prime}\\
-\mathbf{M}_{\rho q}^{f\T\prime}	&	\mathbf{M}_{qq}^{f\T\prime}
\end{pmatrix}\nonumber,
\end{align}
and $\mathbf{M}_{\alpha\beta}^{f\prime}=\frac{\partial\alpha_t^\prime}{\partial\beta_0^\prime}$.

\section{The MInt Algorithm}
\label{ap:alg}
Here we describe the implementation of the MInt algorithm along with 
exact evolution of the Monodromy matrix. 
To avoid computational difficulties with complex numbers the formal 
evolution equations are rewritten such that the algorithm, 
when coded, is entirely real.\\

%%%% POSITION AND MOMENTUM EVOLUTION %%%%
\subsection{Evolution of positions and momenta}
In the following we assume the diabatic electronic potential energy matrix to be real-symmetric, the extension 
to Hermitian $\bigV(\bigR)$
is straightforward.

Evolution of nuclear position is given in \eqr{Rsol}. 

To evolve the electronic positions and momenta in \eqr{xpev}, we diagonalize the diabatic matrix $\bV$ 
giving eigenvectors $\bS$ and a diagonal eigenvalue matrix
 $\bLam$ such that $\bigS\T \bigV \bigS=\bLam$, where we 
drop the $\bR$ dependence of $\bV$, $\bS$, and $\bLam$ for clarity. We then calculate
\begin{subequations}
\begin{align}
\bigC= & \bigS\cos(\bLam\Delta t)\bigS\T \\
\bigD= & \bigS\sin(-\bLam\Delta t)\bigS\T
\end{align} \eql{cd}
\end{subequations}
such that
\begin{subequations}
\begin{align}
\xe(\Delta t)=&\bigC\xe(0)-\bigD\pe(0)\eql{qsol}\\
\pe(\Delta t)=&\bigC\pe(0)+\bigD\xe(0)\eql{psol}.
\end{align} \eql{xpev2}
\end{subequations}
To solve \eqr{nucpev}, we insert $\bS\bS\T = \mathbb{I}$ identities and define 
\begin{align}
\bigW_k := \bigS\T\bigV_k\bigS \eql{wkdef}
\end{align}
to be the derivative of the potential in the adiabatic basis, giving 
\begin{widetext}
\begin{align}
P_k(\Delta t) = P_k(0)-\frac{1}{2} \int_0^{\Delta t} dt \left\{[\xe(0)-i\pe(0)]\T \bS e^{+i\bLam t} \bW_k e^{-i\bLam t}
 \bS\T [\xe(0)+i\pe(0)]- \tr[ \bV_k(\bR)]\right\}. \eql{nucpev2}
\end{align}
\end{widetext}
As defined earlier we use $\bV_k(\bR):=\frac{\partial}{\partial R_k}\bV(\bR)$.
We then integrate the elements of $e^{+i\bLam t} \bW_k e^{-i\bLam t}$ term by term to give
\begin{align}
\int_0^{\Delta t} dt\,e^{+i\bLam t} \bW_k e^{-i\bLam t} = \bm{\Gamma}_k + i\bm{\Xi}_k
\end{align}
where
\begin{subequations}
\begin{align}
(\bigGam_k)_{mn}&=
\begin{cases}
\left(\bigW_{k}\right)_{mn}\Delta t	&	m=n\\
\frac{1}{\lambda_{mn}}\sin(\lambda_{mn}\Delta t)\left(\bigW_k\right)_{mn}	&	m\neq n
\end{cases}\\
(\bigXi_k)_{mn}&=
\begin{cases}
0	&	m=n	\\
\frac{1}{\lambda_{mn}}\left[1-\cos(\lambda_{mn}\Delta t)\right]\left(\bigW_k\right)_{mn}	&	m\neq n\\
\end{cases}
\end{align}\eql{gamxi}%
\end{subequations}
where we use the shorthand $\lambda_{mn}=(\bLam)_{mm}-(\bLam)_{nn}$. Note that $\bigGam_k$ is real 
and symmetric and $\bigXi_k$ is real and skew-symmetric
since by definition $\lambda_{mn}=-\lambda_{nm}$.

We then rotate $\bigGam_k$ and $\bigXi_k$ back to the diabatic basis, defining 
\begin{subequations}
\begin{align}
\bigE_k := & \bigS\bigGam_k\bigS\T, \\
\bigF_k := & \bigS\bigXi_k\bigS\T,
\end{align}\eql{efdef}%
\end{subequations}
where $\bigE$ is symmetric and $\bigF$ is skew-symmetric.
Inserting this into \eqr{nucpev2} we finally obtain
\begin{align}\eql{psolNuc}
P_k(\Delta t)=&P_k(0)-\frac{1}{2}\big\{ \bx\T(0)\bigE_k\bx(0)+\pe\T(0)\bigE_k\pe(0)\nonumber\\
&-2\bx\T(0)\bigF_k\pe(0)-\Tr\left[\bigV_k\right]\Delta t\big\}.
\end{align}

%%%% MONODROMY MATRIX EVOLUTION %%%%
\subsection{Evolution of the monodromy matrix}
From \eqr{mondef}, the monodromy matrix in mapping variables is given as
\begin{align}
\bM = 
\begin{pmatrix}
\bM_{\bR\bR} & \bM_{\bR \bx} & \bM_{\bR \bP} & \bM_{\bR\bp} \\
\bM_{\bx\bR} & \bM_{\bx \bx} & \bM_{\bx \bP} & \bM_{\bx\bp} \\
\bM_{\bP\bR} & \bM_{\bP \bx} & \bM_{\bP \bP} & \bM_{\bP\bp} \\
\bM_{\bp\bR} & \bM_{\bp \bx} & \bM_{\bp \bP} & \bM_{\bp\bp} \\
\end{pmatrix}
\end{align}
where 
\begin{align}
\bM_{\bX\mathbf{Y}} = \ddp{\bX(t)}{\mathbf{Y}(0)}
\end{align}
for two arbitrary phase space variables $\bX$ and $\bY$.

\subsubsection{Evolution under $H_1$}

Since evolution under $H_1$ is linear, for evolution through $\Delta t/2$ the diagonal elements of $\bM$ are unity, 
\begin{align}
\bM_{\bR\bP} = \frac{\Delta t}{2 \mu_{kk}}
\end{align}
and all other elements of $\bM$ are zero. The update to the monodromy matrix is therefore\cite{bre97a}
\begin{align}\eql{RsolM}
\mathbf{M}_{R_k \bX}(\Delta t /2)&=\mathbf{M}_{R_k \bX}(0)+\mathbf{M}_{P_k\bX}(0)\frac{\Delta t}{2\mu_{kk}}
\end{align}
and all other elements are unchanged. 

\subsubsection{Evolution under $H_2$}
We first observe that for the equations of motion in \eqr{h2eq}, $\bM_{\bR\bR} = \bM_{\bP\bP} = \mathbb{I}$, 
and all elements of $\bM_{\bR\bx}$, $\bM_{\bR\bP}$ and $\bM_{\bR\bp}$ are zero.

The monodromy matrix elements concerning only the electronic variables can be obtained from 
\eqr{qsol} and \eqr{psol} at no extra computational cost,
\begin{subequations}
\begin{align}
\bigM_{\bx\bx}(\Delta t)&=\bigC\eql{eM1}\\
\bigM_{\bx\pe}(\Delta t)&=-\bigD\eql{eM2}\\
\bigM_{\pe\bx}(\Delta t)&=\bigD\eql{eM3}\\
\bigM_{\pe\pe}(\Delta t)&=\bigC\eql{eM4}.
\end{align}\eql{Mel}
\end{subequations}
We can similarly use \eqr{psolNuc} to determine changes in nuclear momenta with respect to
initial electronic coordinates,
\begin{subequations}
\begin{align}
\mathbf{M}_{P_k\bx}(\Delta t)&=-\left[\bx\T(0)\bigE_k+\pe\T(0)\bigF_k\right]\\
\mathbf{M}_{P_k\pe}(\Delta t)&=-\left[\pe\T(0)\bigE_k-\bx\T(0)\bigF_k\right].
\end{align}\eql{MPel}%
\end{subequations}
Determining $\bM_{\bx\bR}$ and $\bM_{\bp\bR}$ requires finding the derivative of a matrix exponential. We use \eqr{qsol} and \eqr{psol} to give
\begin{subequations}
\begin{align}
\mathbf{M}_{\bx R_k}(\Delta t)&=\bigC_{k}\bx(0)-\bigD_{k}\pe(0)\eql{mqrk}\\
\mathbf{M}_{\pe R_k}(\Delta t)&=\bigC_{k}\pe(0)+\bigD_{k}\bx(0)\eql{mprk},
\end{align}\eql{MxpR}%
\end{subequations}
where, similar to Appendix A of Ref.~\onlinecite{hel11a}, %$\bigC_k\equiv\frac{\partial}{\partial R_k}\bigC$ and 
\begin{subequations}
\begin{align}
\bigC_k:= & \frac{\partial}{\partial R_k}\bigC \no \\
= & \bS_k \cos(\bLam \Delta t) \bS\T - \bS\sin(\bLam \Delta t)\bLam_k\Delta t \bS\T \no \\ 
& + [\bS_k \cos(\bLam \Delta t) \bS\T]\T, \\
\bigD_k := & \frac{\partial}{\partial R_k}\bigD \no \\
= & -\bS_k \sin(\bLam \Delta t) \bS\T - \bS\cos(\bLam \Delta t)\bLam_k\Delta t \bS\T \no \\ 
& - [\bS_k \sin(\bLam \Delta t) \bS\T]\T, \\
\bS_k := & \ddp{}{R_k} \bS, \\
\bLam_k := & \ddp{}{R_k} \bLam.
\end{align}\eql{CDSL}%
\end{subequations}
For a system with two electronic states $\bS_k$ and $\bLam_k$ can be determined algebraically, and algorithms exist for
 finding these exactly for an arbitrary $F$-level system.\cite{mag85a}

We finally require $\bM_{\bP\bR}$. Differentiating \eqr{psolNuc} gives
\begin{align} %\label{mprnuc}
\bigM_{P_kR_j}=&-\frac{1}{2}\left[\bx\T\mathbf{E}_{jk}\bx+\pe\T\mathbf{E}_{jk}\pe-2\bx\T\mathbf{F}_{jk} \pe\right]\nonumber\\
&+\frac{1}{2}\Tr\left[\bigV_{jk}(\bigR)\right]\Delta t, \eql{MPR}
\end{align}
where 
\begin{subequations}
\begin{align}
\bigV_{jk}:= & \frac{\partial}{\partial R_j}\bigV_k, \\
\mathbf{E}_{jk} := & \ddp{}{R_j} \mathbf{E}_k \no \\
= & \bS_j \bm{\Gamma}_k \bS\T + \bS \bm{\Gamma}_{jk} \bS\T + (\bS_j \bm{\Gamma}_k \bS\T)\T, \\
\mathbf{F}_{jk} := & \ddp{}{R_j} \mathbf{F}_k \no \\
= & \bS_j \bm{\Xi}_k \bS\T + \bS \bm{\Xi}_{jk} \bS\T - (\bS_j \bm{\Xi}_k \bS\T)\T ,
\end{align}\eql{EFjk}
\end{subequations}
and %$\mathbf{\Xi}_{jk}:=\ddp{}{R_j}\mathbf{\Xi}_k$ are given by
\begin{widetext}
\begin{subequations}
\begin{align}
(\mathbf{\Gamma}_{jk})_{mn} := & \ddp{}{R_j}\left(\mathbf{\Gamma}_k\right)_{mn} \no \\
= & \left\{
\begin{array}{ll}
(\bW_{jk})_{nm} \Delta t & m=n \\
\frac{1}{\lambda_{mn}}\sin(\lambda_{mn}\Delta t)\left[(\bW_{jk})_{mn} - \frac{\lambda_{j,mn}}{\lambda_{mn}}(\bW_{k})_{mn}\right] 
+ \frac{1}{\lambda_{mn}}\cos(\lambda_{mn}\Delta t) \lambda_{j,mn}\Delta t (\bW_{k})_{mn} & m \neq n
\end{array}
\right. ,\\
(\mathbf{\Xi}_{jk})_{mn} := & \ddp{}{R_j}\left(\mathbf{\Xi}_k\right)_{mn}\no \\
= & \left\{
\begin{array}{ll}
0 & m=n \\
\frac{1}{\lambda_{mn}}[1-\cos(\lambda_{mn}\Delta t)]\left[(\bW_{jk})_{mn} - \frac{\lambda_{j,mn}}{\lambda_{mn}}(\bW_{k})_{mn}\right] 
+ \frac{1}{\lambda_{mn}}\sin(\lambda_{mn}\Delta t) \lambda_{j,mn}\Delta t (\bW_{k})_{mn} & m \neq n
\end{array}
\right. ,
\end{align}\eql{GamXijk}
\end{subequations}
\end{widetext}
and
\begin{subequations}
\begin{align}
\bW_{jk} := & \ddp{}{R_j}\bW_k \no\\
= & \bS_j\T\bV_k\bS + \bS\T \bV_{jk} \bS + (\bS_j\T\bV_k\bS)\T ,\\
\lambda_{j,mn} := & \ddp{}{R_j} \lambda_{mn} = (\bLam_j)_{mm} - (\bLam_j)_{nn}.
\end{align}\eql{WLjk}
\end{subequations}

Despite the apparent complexity of the monodromy matrix calculations, many terms can be `recycled' from previous 
operations, such as matrices $\bigS$, $\bigC$ and $\bigD$, etc. In addition, for a two-level system $\bS_j \bm{\Xi}_k \bS\T$ 
is diagonal and therefore $\mathbf{F}_{jk} = \bS \bm{\Xi}_{jk} \bS\T$. 

%We have also found that the algorithm allows a 
%far coarser time-step than the Adams propagator \timcom{\texttt{Matt insert numbers here for energy/symp conservation}}
% such that computing a trajectory of given length is cheaper using this algorithm.

\subsection{Complete algorithm}
The trajectory is initialized with given values of $\{\bR,\bx,\bP,\bp\}$ and $\bigM(0)=\mathbb{I}$. Starred items 
are only required if the monodromy matrix is also to be evaluated.

For each time step
\begin{enumerate}
\item Evolve nuclear positions with \eqr{Rsol} for $\Delta t/2$.
\item $^*$Evolve $\bigM$ for $\Delta t/2$ using \eqr{RsolM}.
\item Compute $\bigV$ and $\bigV_k \ \forall \ k$. Diagonalize $\bV$ to find $\bS$ and $\bLam$.
\item Find $\bC$ and $\bD$ using \eqr{cd} and calculate $\bx(t)$ and $\bp(t)$ from \eqr{xpev2}. \label{CD}
\item For each $k$, find $\bW_k$ and from it $\bm{\Gamma}_k$ and $\bm{\Xi}_k$ using \eqr{gamxi}. From 
these obtain $\mathbf{E}_k$ and $\mathbf{F}_k$ $\forall k$ using \eqr{efdef}. Therefore find $\bP(t)$ from \eqr{psolNuc}. \label{EF}
% Monodromy matrix updates
\item $^*$Find $\bV_{jk}$, $\bS_j$, and $\bLam_{jk}$ $\forall \ j,k$. \label{monbegin}
\item $^*$Populate $\bigM_{\bx\bx}$, $\bigM_{\bx\pe}$, $\bigM_{\pe\bx}$, and $\bigM_{\pe\pe}$ from \eqr{Mel} using the $\bC$ and $\bD$ from step~\ref{CD}.
\item $^*$From \eqr{MPel} find $\bM_{\bP\bx}$ and $\bM_{\bP\bx}$ using $\{\mathbf{E}_k\}$ and $\{\mathbf{F}_k\}$ from step~\ref{EF}.
\item $^*$Find $\{\mathbf{C}_k\}$ and $\{\mathbf{D}_k\}$ from \eqr{CDSL} and therefore $\bM_{\bx\bR}$ and $\bM_{\bp\bR}$ from \eqr{MxpR}.
\item $^*$Find $\{\bW_{jk}\}$ and $\{\lambda_{j,mn}\}$ defined in \eqr{WLjk} and compute $\bm{\Gamma}_{jk}$ and $\bm{\Xi}_{jk}$ using \eqr{GamXijk}. 
From these find $\{\mathbf{E}_{jk}\}$ and $\{\mathbf{F}_{jk}\}$ [\eqr{EFjk}] and compute $\bM_{\bP\bR}$ using \eqr{MPR}.\label{monend}
\item $^*$Evolve the monodromy matrix using the monodromy matrix for $\Phi_{H_2,\Delta t}$ obtained from steps~\ref{monbegin} to \ref{monend}.
%\item{Evolve electronic variables for a full time step with Eq.~(\ref{qsol}) and Eq.~(\ref{psol}), and update $\bigM$
%with Eq.s~(\ref{eM1}-\ref{eM4}) as well as with  Eq.~(\ref{mqrk}) and Eq.(\ref{mprk}).}
%\item{Evolve nuclear momenta with Eq.~(\ref{psolNuc}), and update $\bigM$ with Eq.~(\ref{mprnuc}).}
\item{Repeat steps $1$ and $2^*$ for evolution step $\Phi_{H_1,\Delta t/2}$.}
%\item{Repeat steps 2-7 for every additional time step.}
\end{enumerate}

We note that a different flow map constructed by swapping $H_1$ and $H_2$ in \eqr{flowmap}
would also result in a symplectic transformation, but the flow map defined in \eqr{flowmap}
 requires fewer mathematical operations.

\section{Algorithm properties}
A symmetric algorithm is formally defined as\cite{lei04a}
\begin{align}
\Psi_{-\Delta t} = \Psi_{\Delta t}^{-1}.
\end{align}
To prove this, we use the property that exact evolution under any Hamiltonian is symmetric ($\Phi^{-1}_{t} = \Phi_{-t}$) and therefore 
\begin{align}
\Psi^{-1}_{H, \Delta t} = & \Phi_{\Delta t/2, H_1}^{-1} \circ \Phi_{\Delta t, H_2}^{-1} \circ \Phi_{\Delta t/2, H_1}^{-1} \no \\
= & \Phi_{-\Delta t/2, H_1} \circ \Phi_{-\Delta t, H_2} \circ \Phi_{-\Delta t/2, H_1} \no \\
= & \Psi_{H, -\Delta t}
\end{align}
as required. 

Time reversibility is formally\cite{lei04a}
\begin{align}
\Psi_{H,\Delta t} = \bm{\Sigma} \Psi_{H,\Delta t}^{-1}(\bm{\Sigma} \bz)
\end{align}
where the involution $\bm{\Sigma}$ is
\begin{align}
\bm{\Sigma} = \begin{pmatrix} \mathbb{I} & \mathbf{0} \\ \mathbf{0} & -\mathbb{I} \end{pmatrix}.
\end{align}
Exact evolution under the MMST Hamiltonian is time reversible since $H(\bigR,\mathbf{x},\mathbf{P},\mathbf{p})=H(\bigR,\mathbf{x},-\mathbf{P},-\mathbf{p})$. This
can be proven for $\Psi_{H,\Delta t}$ since exact evolution under $H_1$ and $H_2$ is time-reversible and therefore
\begin{align}
\bSig \Psi^{-1}_{H, \Delta t} (\bSig\bz) = & \bSig [\Phi_{\Delta t/2, H_1} \circ \Phi_{\Delta t, H_2} \circ \Phi_{\Delta t/2, H_1}]^{-1}(\bSig\bz)\no\\
= & \bSig \Phi_{\Delta t/2, H_1}^{-1} \circ \Phi_{\Delta t, H_2}^{-1} \circ \Phi_{\Delta t/2, H_1}^{-1}(\bSig\bz) \no\\
= & \bSig \Phi_{\Delta t/2, H_1}^{-1} \circ \Phi_{\Delta t, H_2}^{-1} [\bSig  \Phi_{\Delta t/2, H_1}(\bz)] \no\\
= & \bSig \Phi_{\Delta t/2, H_1}^{-1} [ \bSig \Phi_{\Delta t, H_2} \circ \Phi_{\Delta t/2, H_1}(\bz)] \no\\
= & \Phi_{\Delta t/2, H_1} \circ \Phi_{\Delta t, H_2} \circ \Phi_{\Delta t/2, H_1}(\bz)\no \\
= & \Psi_{\Delta t}(\bz).
\end{align}
To show that the algorithm is second order, one can write out exact evolution under $H$ in powers of $\Delta t$ using the 
Liouvillian formalism and then compare to evolution under $\Psi_{H,\Delta t}$, noting that terms differ at $\mathcal{O}(\Delta t^3)$. 
More elegantly, since a method constructed by Hamiltonian splitting is exactly symplectic and at least first order,\cite{lei04a} and that 
a symmetric method has to be of even order,\cite{lei04a} the algorithm must be (at least) second order accurate.

To prove that $\mathcal{G}:= \bx\T\bx + \bp\T\bp$ is conserved, we note that it is unchanged by evolution under 
$H_1$, i.e.\ $\{\mathcal{G},H_1\} = 0$ and for evolution under $H_2$ we find $\{\mathcal{G},H_2\} = 2\bx\T\bV\bp - 2\bp\T\bV\bx = 0$ as $\bV$ is symmetric.

Angle invariance is a direct consequence of unitarity.\cite{mey79b} To show this explicitly one can apply the transformation in \eqr{theta} to \eqr{xpev} and then transform back, observing 
that evolution of the electronic positions and momenta are unaffected. The evolution of nuclear position in \eqr{Rsol} is 
not directly dependent on the electronic variables and evolution of nuclear momenta in \eqr{nucpev} is invariant to the transformation in \eqr{theta}.

Since the MInt algorithm is Hamiltonian evolution discretized by a symplectic method, there exists a \emph{modified Hamiltonian} $\check H$ 
whose energy the algorithm conserves exponentially well over exponentially long time intervals\cite{lei04a}. The modified Hamiltonian, 
which is timestep-dependent, differs from the original Hamiltonian by the order of the algorithm,\cite{lei04a} so for the MInt algorithm
\begin{align}
H(\bz) - \check H(\bz;\Delta t) = \mathcal{O}(\Delta t^2)
\end{align}
and the MMST Hamiltonian $H(\bz)$ will be conserved for exponentially long times with fluctuations of $\mathcal{O}(\Delta t^2)$. 

\section{Liouvillian formalism}
\label{ap:liou}
The algorithm in \eqr{flowmap} in the Liouvillian representation is equivalent to
\begin{align}
\Psi_{H,\Delta t} = e^{\mL_1 \Delta t/2}e^{\mL_2 \Delta t} e^{\mL_1 \Delta t/2} \eql{liou}
\end{align}
where
\begin{subequations}
\begin{align}
\mL_1 = & \{\cdot,H_1\} \no \\
= & \sum_k \frac{P_k}{\mu_{kk}} \ddp{}{R_k}, \eql{l1} \\
\mL_2 = & \{\cdot,H_2\} \no \\
= & -\sum_k \Bigg\{ \frac{1}{2}(\xe-i\pe)\T\bigV_k(\bigR)(\xe+i\pe)\no \\
& \qquad -\frac{1}{2}\Tr\left[\bigV_k(\bigR)\right] \Bigg\} \ddp{}{P_k} \no\\
& + \bp\T \bV \nabla_\bx - \bx\T\bV\nabla_\bp. \eql{l2}
\end{align} \eql{ll}%
\end{subequations}
%, i.e.\ 
%\begin{align}
%\{\cdot,H\} \equiv - (\nabla_\bz H)\T \bJ \nabla_\bz
%\end{align}
Note that each Liouvillian can be written as exact evolution under a Hamiltonian, and we follow the conventions 
of Zwanzig\cite{zwa01a} and Ref.~\onlinecite{lei04a} by defining the Liouvillian without a prefactor of $i$.

An alternative scheme has been suggested for evolution in mapping variables which (in this notation) is\cite{ric16b}
\begin{align}
\tilde \Psi_{H,\Delta t} = e^{\mL_{\rm el} \Delta t/2} e^{\mL_{\bP} \Delta t/2} e^{\mL_1 \Delta t}  e^{\mL_{\bP} \Delta t/2} e^{\mL_{\rm el} \Delta t/2} \eql{tildpsi}
\end{align}
where $\mL_1$ is defined in \eqr{l1} and
\begin{subequations}
\begin{align}
\mL_{\rm el} = & + \bp\T \bV \nabla_\bx - \bx\T\bV\nabla_\bp \\
\mL_{\bP} = & -\sum_k \Bigg\{ \frac{1}{2}(\xe-i\pe)\T\bigV_k(\bigR)(\xe+i\pe)\no \\
& \qquad -\frac{1}{2}\Tr\left[\bigV_k(\bigR)\right] \Bigg\} \ddp{}{P_k}.
\end{align}\eql{tilde}
\end{subequations}

To compare these algorithms, we firstly note that the order of $\mL_1$ and $\mL_2$ in \eqr{liou} can be 
swapped without compromising the formal properties of the algorithm
\begin{align}
\bar \Psi_{H,\Delta t} = e^{\mL_2 \Delta t/2}e^{\mL_1 \Delta t} e^{\mL_2 \Delta t/2} \eql{liou}
\end{align}
though this will then be more computationally expensive than $\Psi_{H,\Delta t}$. We then note from \eqr{l2} and \eqr{tilde} that 
\begin{align}
\mL_2 \equiv \mL_{\rm el} + \mL_{\bP}.
\end{align}
Consequently $\tilde \Psi_{H,\Delta t}$ is equivalent to making the approximation
\begin{align}
e^{\mL_2 \Delta t/2} \simeq e^{\mL_{\rm el} \Delta t/2} e^{\mL_{\bP} \Delta t/2} \eql{approx}
\end{align}
to the symplectic propagator $\bar \Psi_{H,\Delta t}$. We therefore call $\bar \Psi_{H,\Delta t}$ the Split 
Liouvillian (SL) algorithm since it splits $e^{\mL_2 \Delta t/2}$ into $e^{\mL_{\rm el} \Delta t/2} e^{\mL_{\bP} \Delta t/2}$ 
(and $e^{\mL_{\bP} \Delta t/2} e^{\mL_{\rm el} \Delta t/2}$). 

The approximation in \eqr{approx} is clearly exact in the $\Delta t \to 0$ limit, and therefore $\tilde \Psi_{H,\Delta t}$ 
will be symplectic in this limit. It will also conserve electronic probability exactly for any time step like $\bar \Psi_{H,\Delta t}$ and $ \Psi_{H,\Delta t}$.

However, $\mL_{\rm el}$ and $\mL_{\bP}$ cannot in general be written as exact evolution under a Hamiltonian [cf.~\eqr{ll}] 
and we show in appendix~\ref{ap:symp} that the SL algorithm is not in general symplectic for an arbitrary timestep.

\section{Symplecticity properties of the MInt and SL algorithms}
\label{ap:symp}
Here we confirm that the MInt algorithm is symplectic by explicitly evaluating \eqr{sympcrit} for each step of the algorithm. 
We also show that the SL algorithm in \eqr{tildpsi} is not, in general, symplectic. For notational simplicity we present the 
results for one nuclear dof; further nuclear dof merely add more indices.

We first note that evolution under an arbitrary series of symplectic steps is also symplectic, since the monodromy 
matrix of the overall algorithm is the product of the monodromy matrices of the individual steps, and symplecticity can 
therefore be proven by applying \eqr{sympcrit} recursively. To prove that the MInt algorithm is symplectic it is therefore sufficient to prove
\begin{align}
\bM_{H_1}\T\bJ^{-1}\bM_{H_1} = \bJ^{-1} \eql{h1crit}
\end{align}
and
\begin{align}
\bM_{H_2}\T\bJ^{-1}\bM_{H_2} = \bJ^{-1} \eql{h2crit}
\end{align}
where $\bM_{H_1}$ and $\bM_{H_2}$ are the monodromy matrices associated with evolution under $H_1$ and $H_2$ respectively.

\subsection{Evolution under $H_1$}
\label{ssec:h1ev}
The monodromy matrix (for evolution with timestep $\Delta t/2$) is simply
\begin{align}
\bM_{H_1} = 
\begin{pmatrix}
1 & \mathbf{0}\T & \Delta t/2m & \mathbf{0}\T \\
\bzero & \mathbb{I} & \bzero & \mathbb{O} \\
0 & \bzero\T & 1 & \bzero\T \\
\bzero & \mathbb{O} & \bzero & \mathbb{I}
\end{pmatrix}
\end{align}
where $\bzero$ is the null vector. Simple matrix multiplication shows that this satisfies \eqr{h1crit}.

\subsection{Evolution under $H_2$}
\label{ssec:h2ev}
\newcommand{\sfa}{\mathsf{a}}
\newcommand{\sfb}{\mathsf{b}}
\newcommand{\sfe}{\mathsf{e}}
\newcommand{\sff}{\mathsf{f}}
\newcommand{\sfg}{\mathsf{g}}
\newcommand{\sfh}{\mathsf{h}}
\newcommand{\sfj}{\mathsf{j}}
\newcommand{\sfA}{\mathsf{A}}
\newcommand{\sfB}{\mathsf{B}}
\newcommand{\bE}{\mathbf{E}}
\newcommand{\bF}{\mathbf{F}}
\newcommand{\bGam}{\bm{\Gamma}}
%\timcom{Clarify notation with $\bone,\bzero,\mathbb{I}$. Should we use different notation for vectors and matrices?}

We firstly define
\begin{subequations}
\begin{align}
\sfa  = & -\bp\T \mathbf{E} + \bx\T\mathbf{F} \\
\sfb = & -\frac{1}{2}\left[\bx\T\mathbf{E}'\bx+\pe\T\mathbf{E}'\pe-2\bx\T\mathbf{F}' \pe\right]\nonumber\\
&+\frac{1}{2}\Tr\left[\bigV''\right]\Delta t \\
\sfe = & -\bx\T\mathbf{E} - \bp\T\mathbf{F} \\
\sff  = & \bC'\bp + \bD'\bx \\
\sfg = & \bC'\bx - \bD'\bp 
\end{align}\eql{sfdefs}%
\end{subequations}
where the primes denote derivatives w.r.t.\ the nuclear co-ordinate, such that
\begin{align}
\bM_{H_2} = 
\begin{pmatrix}
1 & \mathbf{0}\T & 0 & \mathbf{0}\T \\
\mathsf{g} & \mathbf{C} & \mathbf{0} & -\mathbf{D}  \\
\mathsf{b} & \mathsf{e} & 1 & \mathsf{a} \\
\mathsf{f} & \mathbf{D} & \mathbf{0} & \mathbf{C}  \\
\end{pmatrix}\eql{mh2}
\end{align}
and
\begin{widetext}
\begin{align}
\bM_{H_2}\T \bJ^{-1} \bM_{H_2} = 
\begin{pmatrix}
0& -\sfe - \sfg\T\bD + \sff\T\bC & -1 & -\sfa - \sfg\T\bC - \sff\T\bD \\
-\bC\sff + \sfe\T + \bD \sfg & -\bC\bD + \bD\bC & \bzero & -\bC\bC - \bD\bD  \\
1 & \bzero & 0 & \bzero & \\
\bD\sff + \sfa\T + \bC\sfg & +\bD\bD + \bC\bC & \bzero & +\bD\bC - \bC\bD
\end{pmatrix}. \eql{mjm2}
\end{align}
\end{widetext}%\timcom{check this}
We firstly note that $\bC\bD - \bD\bC = \mathbb{O}$ since these matrices have the same eigenvectors and $\bC\bC + \bD\bD = \mathbb{I}$. We then define
\begin{subequations}
\begin{align}
\sfh := & -\bC\sff + \sfe\T + \bD \sfg \\
\sfj := & \bD\sff + \sfa\T + \bC\sfg ,
\end{align}\eql{hj}%
\end{subequations}
such that \eqr{mjm2} reduces to
\begin{align}
\bM_{H_2}\T \bJ^{-1} \bM_{H_2} = 
\begin{pmatrix}
0 & -\sfh\T & -1 & -\sfj\T \\
\sfh & \mathbb{O} & \bzero & -\mathbb{I} \\
1 & \bzero & 0 & \bzero \\
\sfj & \mathbb{I} & \bzero & \mathbb{O}
\end{pmatrix}. \eql{MJM}
\end{align}
To evaluate \eqr{hj} we define the matrices
\begin{subequations}
\begin{align}
\sfA := & \bD\bC' - \bE - \bC\bD' \eql{cond1} \\
\sfB := & -(\bD\bD' - \bF + \bC\bC') \eql{cond2},
\end{align}
\end{subequations}
such that
\begin{subequations}
\begin{align}
\sfh \equiv & \sfA \bx + \sfB \bp \eql{hab} \\
\sfj \equiv & - \sfB \bx + \sfA \bp. \eql{jba}
\end{align}
\end{subequations}%\timcom{check this}
In order to prove \eqr{h2crit}, we must prove $\sfh \equiv \mathbf{0}$ and $\sfj \equiv \mathbf{0}$ $\forall$ $\bx,\bp$, which 
requires proving $\sfA \equiv \mathbb{O}$ and $\sfB \equiv \mathbb{O}$. As we shall see, it is mathematically convenient to 
prove this in the adiabatic basis, i.e.\ $\bS\T\sfA\bS \equiv \mathbb{O}$ and $\bS\T\sfB\bS \equiv \mathbb{O}$.

We find
\begin{align}
\bS\T\sfA\bS = & \bLam' t - \sin(\bLam\Delta t)\bS\T\bS'\cos(\bLam \Delta t) \no\\
& + \cos(\bLam\Delta t)\bS\T\bS'\sin(\bLam\Delta t) - \bGam
\end{align}
such that
\begin{align}
(\bS\T\sfA\bS)_{nm} = \bLam'_{nn} \delta_{nm} \Delta t + (\bS\T\bS')_{nm}\sin(\lambda_{mn} \Delta t) - \bGam_{nm} \eql{cond}
\end{align}
To evaluate the $\bW$ matrix in $\bGam$, we find from \eqr{wkdef}
\begin{align}
\bW %= &  \bS\T\left(\dd{}{R} \bV \right)\bS \\
= & \bS\T\left(\ddp{}{R} \bS \bLam \bS\T \right)\bS \no \\
= & \bS\T\bS' \bLam + \bLam' + \bLam \mathbf{S}'^{\mathrm T} \bS.
\end{align}
We also use the property that the nonadiabatic derivative coupling matrix $\bS\T\bS'$ is 
antisymmetric, i.e.\ because $\bS\T\bS = \mathbb{I}$, $\bS'^{\mathrm{T}} \bS + \bS\T\bS' = \mathbb{O}$, and therefore
\begin{align}
\bW_{nm} = & (\bS\T\bS')_{nm} \lambda_{mn} + \bLam'_{nn}\delta_{nm}.
\end{align}
Inserting this into \eqr{gamxi} we obtain
\begin{subequations}
\begin{align}
\bGam_{nm} =  & 
 \left\{
 \begin{array}{cl}
 \bLam'_{nn} \Delta t & n=m \\
-(\bS\T\bS')_{nm}\sin(\lambda_{nm}\Delta t) & n \neq m 
 \end{array}
 \right. \eql{newgam} \\ 
\bm{\Xi}_{nm} = & 
 \left\{
 \begin{array}{cl}
 0 & n=m \\
 \left[\cos( \lambda_{nm} \Delta t ) - 1\right] (\bS\T \bS' )_{nm} & n \neq m 
 \end{array}
 \right. . \eql{newxi}
\end{align}
\end{subequations}
Inserting \eqr{newgam} into \eqr{cond} shows that $\bS\T\sfA\bS \equiv \mathbb{O}$ and therefore $\sfA \equiv \mathbb{O}$. 

To prove that $\sfB = \mathbb{O}$, we find
\begin{align}
(\bS\T\sfB\bS)_{nm} = -(\bS\T\bS')_{nm}[\cos(\lambda_{nm} \Delta t)-1] + \bm{\Xi}_{nm}
\end{align}
since $\bS\T\bS$ is skew-symmetric (see above) then the diagonal elements of this will vanish, and the 
off-diagonal elements also vanish by \eqr{newxi}, such that $\sfB \equiv \mathbb{O}$. Consequently $\sfh = \mathbf{0}$ 
by \eqr{hab} and $\sfj = \mathbf{0}$ by \eqr{jba}, proving that evolution under $H_2$ is symplectic. Combining 
this with section~\ref{ssec:h1ev} proves that $\Psi_{H,\Delta t}$ (the MInt algorithm) and $\bar \Psi_{H,\Delta t}$ 
are symplectic for any timestep, confirming our earlier statement of symplecticity which was based upon contructing 
a method by Hamiltonian splitting\cite{lei04a}.

\subsection{The SL algorithm} %\timcom{Should we change to have a $\Delta t/2$ timestep for $\mL_{\bP}$ and $\mL_{\rm el}$ evolution?}
As noted above, the only difference between $\bar \Psi_{H,\Delta t}$ (which we have just proven to be symplectic) 
and the SL algorithm $\tilde \Psi_{H,\Delta t}$ is the approximation in \eqr{approx}. We therefore seek to determine 
whether successive evolution under $\mL_{\rm el}$ then $\mL_{\bP}$ is symplectic.
The monodromy matrix associated with nuclear momentum evolution (for timestep $\Delta t$) is
\begin{align}
\bM_{\bP} = 
\begin{pmatrix}
1 & \mathbf{0}\T  & 0 & \mathbf{0}\T  \\
\mathbf{0} & \mathbf{1} & \mathbf{0} & \mathbf{0}  \\
 \tilde{\sfb} & - \bq\T\bV' \Delta t & 1 & - \bp\T \bV' \Delta t \\
\mathbf{0} & \mathbf{0} & \mathbf{0} & \mathbf{1} 
\end{pmatrix}
\end{align}
and the matrix associated with electronic evolution only is 
\begin{align}
\bM_{\rm el} = 
\begin{pmatrix}
1 & \mathbf{0}\T & 0 & \mathbf{0}\T \\
\sfg & \mathbf{C} & \mathbf{0} & -\mathbf{D} \\
0 &\mathbf{0}\T & 1 & \mathbf{0}\T \\
\sff & \mathbf{D} & \mathbf{0} & \mathbf{C} 
\end{pmatrix}
\end{align}
where $ \sff$ and $\sfg$ are defined in \eqr{sfdefs} and
\begin{align}
\tilde{\sfb}:=-\frac{1}{2}\left(\xe\T \bV''\xe+\pe\T \bV''\pe-\Tr\left[\bV''\right]\right).
\end{align}

We firstly note that $\det | \bM_{\bP}| \equiv 1$ and $\det | \bM_{\rm el} | \equiv 1$, which means that the 
SL algorithm will satisfy Liouville's theorem, a necessary but not sufficient criterion for symplecticity.

However,
\begin{align}
\bM_{\bP}\T \bJ^{-1} \bM_{\bP} = 
\begin{pmatrix}
0 & \bx\T \bV'\Delta t & -1 & \bp\T\bV' \Delta t \\
-\bV'\bx \Delta t & \mathbb{O} & \bzero & -\mathbb{I} \\
1 & \bzero\T & 0 & \bzero\T & \\
-\bV' \bp \Delta t & \mathbb{I} & \bzero & \mathbb{O}
\end{pmatrix}\eql{pjp}
\end{align} %\timcom{check this}
so evolution under $\mL_{\bP}$ is not symplectic unless $\bV' = 0$ (the diabatic matrix has no nuclear dependence). Furthermore,
\begin{align}
& \bM_{\rm el}\T\bJ^{-1}\bM_{\rm el}  \no\\
=& \begin{pmatrix}
0 & -\sfg\T\bD + \sff\T\bC & -1 & -\sfg\T\bC - \sff\T\bD \\
-\bC\sff + \bD\sfg & \bzero & \bzero & -\mathbb{I} \\
\bone & \bzero & 0 & \bzero \\
\bD\sff + \bC\sfg & \mathbb{I} & \bzero & \bzero
\end{pmatrix} \nonumber\\
\equiv & \begin{pmatrix}
0 & \sfe & -1 & \sfa \\
-\sfe\T & \bzero & \bzero &-\mathbb{I} \\
\bone & \bzero & 0 & \bzero \\
-\sfa\T & \mathbb{I} & \bzero & \bzero 
\end{pmatrix} \eql{eje}
\end{align}%\timcom{check this}
where we have exploited \eqr{hj} and the earlier proofs that $\sfh \equiv \bzero$ and $\sfj \equiv \bzero$. In 
general $\sfa \neq \bzero$ and $\sfe \neq \bzero$, so evolution under $\mL_{\rm el}$ is not symplectic. 

We also consider combined evolution of both $\mL_{\bP}$ and $\mL_{\rm el}$ in order to compare the SL and 
MInt algorithms on an equal footing and show that the combination of steps does not lead to cancellation of errors 
which restores symplecticity. We consider evolution under $\mL_{\bP}$ followed by $\mL_{\rm el}$ (the fourth and 
fifth steps of the SL algorithm), since evolution under $\mL_{\bP}$ first does not change the electronic dofs
 subsequently used in $\bM_{\rm el}$ and therefore leads to simpler algebra. We find
\begin{align}
\bM_{\rm el}\bM_{\bP} = 
\begin{pmatrix}
1 & \bzero & 0 & \bzero \\
\sfg & \bC & \bzero & -\bD \\
\tilde\sfb & -\bx\T\bV' \Delta t & 1 & -\bp\T\bV' \Delta t \\
\sff & -\bD & \bzero & \bC
\end{pmatrix}
\end{align}%\timcom{check this}
comparison with \eqr{mh2} leads us to define
\begin{subequations}
\begin{align}
\tilde \sfa := & -\bp\T\bV' \Delta t \\
\tilde \sfe := & -\bx\T\bV'\Delta t
\end{align}
\end{subequations}
such that
\begin{align}
\bM_{\rm el}\bM_{\bP} = 
\begin{pmatrix}
1 & \bzero & 0 & \bzero \\
\sfg & \bC & \bzero & -\bD \\
\tilde\sfb & \ti\sfe & 1 & \ti \sfa \\
\sff & -\bD & \bzero & \bC
\end{pmatrix}
\end{align}
Comparison with section~\ref{ssec:h2ev} means that $\bM_{\bP}\T \bM_{\rm el}\T \bJ^{-1} \bM_{\rm el}\bM_{\bP} = \bJ^{-1}$ 
if and only if $\ti \sfa \equiv \sfa$ and $\ti \sfe \equiv \sfe$, since the $\tilde\sfb$ term cancels out. Expanding these conditions in coefficients of $\bx$ and $\bp$ 
leads to the conditions %\timcom{not sure that this is the best notation}
\begin{subequations}
\begin{align}
\bE \overset{?}{=} & \bV' \Delta t \\
\bF \overset{?}{=} & \bzero.
\end{align}
\end{subequations}
Evaluating these in the adiabatic basis (as above) gives
\begin{subequations}
\begin{align}
\bS\T(\bE - \bV' \Delta t)\bS = & \bGam - \bW\Delta t \\
\bS\T \bF \bS = & \bm{\Xi} 
\end{align}
\end{subequations}
and evaluating these elementwise in powers of $\Delta t$ gives
\begin{subequations}
\begin{align}
(\bGam - \bW\Delta t)_{nm} = & \left\{
\begin{array}{ll}
0 & n=m \\
%\left[\frac{1}{\Lambda_{nm}} \sin(\Lambda_{nm}\Delta t) - \Delta t\right]W_{nm} & n\neq m
-\frac{\lambda_{nm}^2}{3!} \Delta t^3 W_{nm} + \mathcal{O}(\Delta t^5) & n\neq m
\end{array}
\right.
\\
\bm{\Xi}_{nm} = & \left\{
\begin{array}{ll}
0 & n=m \\
\frac{\lambda_{nm}}{2!}\Delta t^2 W_{nm}  + \mathcal{O}(\Delta t^4) & n\neq m
\end{array}
\right.
\end{align} 
\end{subequations}%\timcom{check this, particularly the signs of terms, $\Lambda_{nm}$ or $\Lambda_{mn}$}
This means that $(\ti \sfe-\sfe)$ and $(\ti \sfa-\sfa)$ will be $\mathcal{O}(\Delta t^2)$. The SL algorithm will therefore be symplectic 
in the $\Delta t \to 0$ limit (as noted above) but for an arbitrary timestep will not be symplectic. Consequently the 
energy is likely to drift, though the extent of the drift may be small if the adiabatic states are closely separated and there 
is little off-diagonal coupling in the adiabatic basis (i.e.\ $\lambda_{nm}W_{nm}\Delta t^2 \ll 1$). We also observe that the 
combination $\bM_{\rm el}\bM_{\bP}$ is symplectic to one higher order in time to $\bM_{\rm el}$ or $\bM_{\bP}$ which 
from \eqr{pjp} and \eqr{eje} will be symplectic to $\mathcal{O}(\Delta t)$.

\bibliography{mybib}
\bibliographystyle{tim}

\end{document}